# High-yield fabrication of bubble-free magic-angle twisted bilayer graphene devices with high twist-angle homogeneity

J. Díez-Mérida[1,2,3], I. Das[2,3], G. Di Battista[2,3], A. Díez-Carlón[2,3], M. Lee[2,3], L. Zeng[4], K. Watanabe[5], T. Taniguchi[5], E. Olsson[4] and D. K. Efetov[2,3]*

1. ICFO - Institut de Ciencies Fotoniques, The Barcelona Institute of Science and Technology, Castelldefels, Barcelona, 08860, Spain
2. Fakultät für Physik, Ludwig-Maximilians-Universität, Schellingstrasse 4, 80799 München, Germany
3. Munich Center for Quantum Science and Technology (MCQST), München, Germany
4. Department of Physics, Chalmers University of Technology, Gothenburg SE-41296, Sweden
5. National Institute of Material Sciences, 1-1 Namiki, Tsukuba, 305-0044, Japan

*E-mail: dmitri.efetov@lmu.de

**Magic-angle twisted bilayer graphene (MATBG) stands as one of the most versatile materials in condensed-matter physics due to its hosting of a wide variety of exotic phases while also offering convenient tunability. However, the fabrication of MATBG is still manual, and remains to be a challenging and inefficient process, with devices being highly dependent on specific fabrication methods, that often result in inconsistency and variability. In this work, we present an optimized protocol for the fabrication of MATBG samples, for which we use deterministic graphene anchoring to stabilize the twist-angle, and a careful bubble removal techniques to ensure a high twist-angle homogeneity. We use low-temperature transport experiments to extract the average twist-angle between pairs of leads. We find that up to ~ 38% of the so fabricated devices show μm² sized regions with a twist-angle in the range $\theta = 1.1 \pm 0.1°$, and a twist-angle variation of only $\Delta\theta \leq 0.02°$, where in some instances such regions were up to 36 μm² large. We are certain that the discussed protocols can be directly transferred to non-graphene materials, and will be useful for the growing field of moiré materials.**

## 1.Introduction

The electronic flat-bands in magic-angle twisted bilayer graphene (MATBG) have shown a rich abundance of emergent quantum phases, such as correlated insulators (CIs)[1–3], superconductors (SCs)[2–4], magnets [5–7], non-trivial topological[8–11] and strange metal phases[12–14]. Similar phases have been unraveled in other moiré materials, such as in twisted bilayers of transition metal dichalcogenides (TMDs)[15,16] and twisted mirror symmetric graphene multi-layers [17–19]. While extensive efforts have been dedicated to understanding the intricate ground states of these systems and what drives them, the community is still struggling to grasp the full details of its colorful phase diagrams. One big challenge endures – device fabrication and sample quality. Fabrication of moiré materials remains notoriously tedious and low yield, and devices are quite sensitive to the details of the fabrication protocols showing strong inhomogeneity and irreproducibility.

The complexity of these rich and diverse phase spaces is influenced by numerous external factors. These factors include the twist-angle[20], the twist-angle disorder[21], the dielectric environment[22,23], the relative alignment to the encapsulating layer used, in particular hexagonal boron nitride (hBN)[6], and strain[21,24,25]. Some of these parameters, such as the selection of specific dielectric thickness ranges or controlling the alignment of graphene with the hBN, are integrated into the fabrication process. Others, like strain or angle disorder, are currently difficult to control and are highly dependent on the stacking process. Therefore, optimizing and standardizing the fabrication process of moiré materials can greatly impact the reproducibility and overall understanding of the intrinsic properties of these materials.

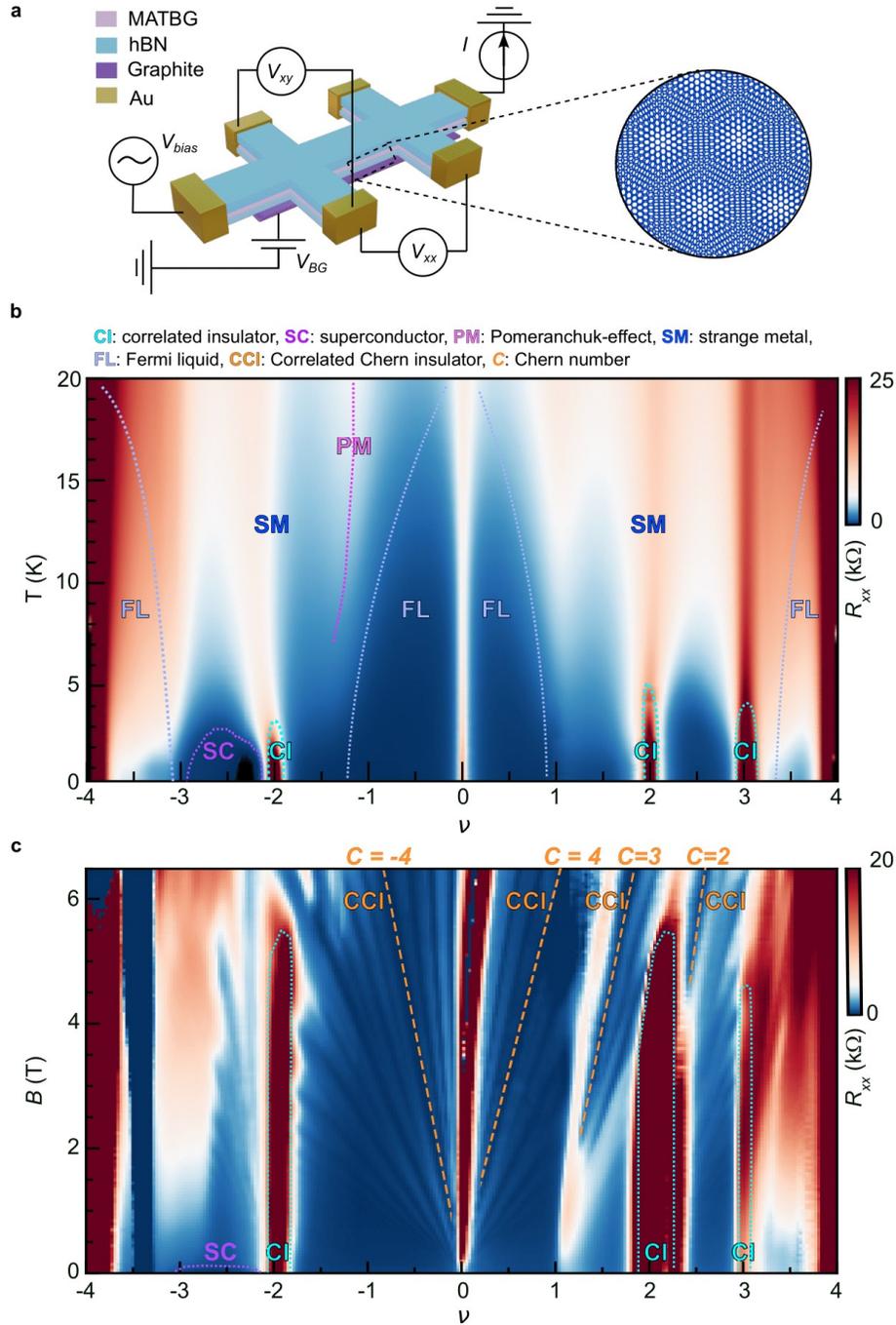

**Fig. 1. Typical low-temperature phase-diagrams of high-quality MATBG devices. a,** Schematic of a hBN encapsulated magic-angle twisted bilayer graphene Hall bar with the typical measuring circuit. The zoom-in shows the moiré pattern formed by twisting the two graphene layers. **b,** $R_{xx}$ vs. filling factor $\nu$ vs. temperature $T$ of a $\theta = 1.16°$ MATBG device displaying CI states at $\nu = \pm 2$ and SC at $\nu = -2 - \delta$. The dashed lines schematically mark the different correlated states, described above the figure. **c,** Landau-fan map ($R_{xx}$ vs. $\nu$ magnetic field $B$) of a $\theta = 1.12°$ MATBG device. The device displays Landau fans emerging from the CI states at $\nu = +1, \pm 2$ and $+3$ and a SC dome at $\nu = -2 - \delta$. The dashed orange lines mark the correlated Chern insulators emerging at each integer filling.

Here, we report a detailed fabrication protocol that was optimized for the high-yield and high-quality assembly of MATBG devices. It is based on a modified dry-transfer technique[26–28], that allows to produce almost bubble free MATBG devices and results in high twist-angle precision and high twist-angle homogeneity. Fig. 1a show the device cross section of a typical so-fabricated MATBG Hall bar device, and Fig. 1b-c shows typical high-quality phase-diagrams of the low temperature transport measurements of the longitudinal resistance $R_{xx}$ vs. the electron filling factor per moiré unit cell $\nu$, as a function of temperature $T$ for a $\theta = 1.16°$ device (b) and $R_{xx}$ vs. $\nu$ and perpendicular magnetic field $B$ for for a $\theta = 1.12°$ device (c). These reveal a high twist-angle homogeneity and rich feature size, and show most of the previously reported phases, such as correlated insulators[1–3], superconductors[2–4], strange metals[12,14], Chern insulators[8–11] and the Pomeranchuk effect[29,30].

## 2. Preparation of the 2D crystals

One of the often overlooked, but key steps, is the careful preparation and selection of appropriate 2D crystals, from which the MATBG is eventually assembled. This determines to a big part the successful outcome. We find that only with properly selected and prepared 2D crystals one can achieve a high yield and high homogeneity of the final stack. Further we discuss in great detail our preparation and selection criteria.

**Exfoliation -** The 2D crystals, in particular the graphene and graphite flakes, are exfoliated via the scotch tape technique on Si/SiO$_2$ (285 nm) substrates, following the standard recipe that was developed by Huang *et al.*[31]: pre-cleaning the chips in O$_2$ plasma and heating up the substrate to ~100 °C for ~2 min to increase the exfoliation yield. For the hBN crystals we however use a slightly altered recipe, where first, we prepare a second or "daughter" scotch tape with thinner hBN crystals, by directly peeling the original tape. And second, we do not apply heat prior to the peeling process, as the hBN tape is very sparsely covered with crystals, as compared to the graphene tape, and will leave too much tape glue residues on the chips (see Suppl. Section A for more details).

For double-graphite-gated devices, the top graphite gate is exfoliated without performing O$_2$ plasma cleaning. This reduces the density of viable flakes which are left on the SiO$_2$, but significantly improves the pick-up probability of the flake of interest, as well as the smoothness of the process, as has been previously reported[26,32]. The bottom graphite gates are produced as a byproduct of graphene exfoliation and can be selected from the plasma cleaned chips.

**Flake selection -** After the exfoliation process, the 2D flakes are screened under the optical microscope and all highly suitable flakes are identified and catalogued. There are several considerations for choosing the individual flakes as well as the relations between the different flakes of the stack (displayed in Fig. 2). In general, the first order criterium in identifying viable flakes is how pristine and homogeneous they are. Selected flakes should have no tape residues, nor step-terraces and should be well isolated from nearby bulky flakes which typically cause problems during the stacking process. Then, there are certain constraints to consider for the different materials.

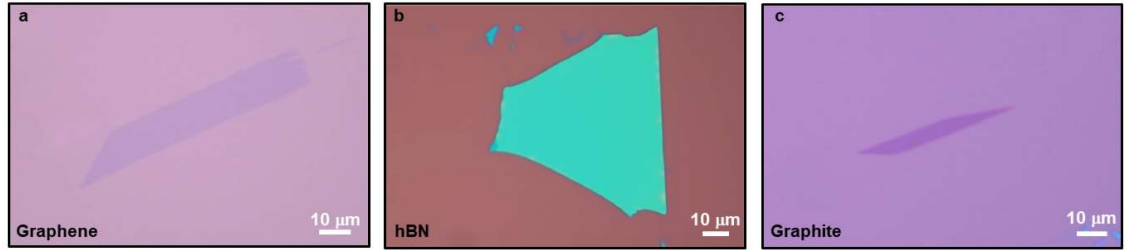

**Fig 2. Graphene and hBN flake selection. a,** A suitable graphene flake has a large size and is not surrounded by thicker flakes. **b,** A suitable hBN flake has a homogenous thickness of 10 - 20 nm, is not surrounded by thicker flakes, and has a sharp edge for clamping. **c,** A suitable graphite gate is straight, 10 μm long, 3 - 4 μm wide and around 2 nm thick.

Graphene flakes – these should be at least twice as large as the desired device size. Flakes which are ~ 10 - 15 μm x 15 - 30 μm are typically desired, such that the final Hall bar devices are ~ 10 μm long.

hBN flakes - should fully encapsulate the graphene and are chosen to be 10 - 20 nm thick, which is considerably thinner than the typically used 30 – 70 nm thick hBN typically used in the community[3–5]. We find that the thinner hBN has several advantages. Thinner hBN flakes are considerably more elastic than thicker flakes, which makes the stacking process much smoother and may generally help in the strain relaxation within the hetero-structure. It also aids to avoid unexpected rapid movements or "jumps" in the stamp during the stacking, which can give rise to sudden stress release and so enhance bubble formation [33] (an example is shown in Suppl. Video 1), which we try minimize as much as possible since these significantly contribute to angle inhomogeneity[21]. However, if the flakes are too thin (i.e. below 5 nm) they are structurally weak and may tear during the stacking process. Also unwanted tunneling or capacitive coupling to the gate electrode may affect the final device[34]. It is also easier to spot dirt, defects, folds or wrinkles in thin hBN flakes under the optical microscope, and flakes below 20 nm thickness are transparent enough to see through them during the stacking process, which is particularly helpful when making multilayered stacks.

The hBN flakes are generally chosen such that they are larger than the graphite gates, to prevent the graphite and graphene layers from shorting in the stack. Also, the hBN which will be picked up first should have at least one sharply defined edge. As we will explain latter, this edge can be used as an anchoring line for the graphene sheets in the pick-up process, which helps stabilizing and locking the crystallographic orientation of the graphene sheets in the hetero-structure.

Graphite flakes – the graphite flakes for the gate electrodes are chosen to be ~ 2 - 4 nm thick, 3 - 6 μm wide and 10 - 15 μm long. The width is chosen such that in the final transport devices the arms of the Hall bars, which extend beyond the width of the device, can be gated away from the charge neutrality point using the highly doped Si substrate, which helps minimize the contact resistance. Flakes below four layers are avoided due to their potentially complex properties, including magnetism in rhombohedral trilayer graphene[35] and their insufficient screening of the charge puddles in the $SiO_2$ substrate[36]. Thicker flakes are also avoided since they are less elastic and may induce more strain to the final stack, and since they are narrower than the twisted graphene regions, they also produce an unwanted height step and curvature in the TBG device, which is directly proportional to the graphite thickness [37]. The bottom gate needs to be longer than graphene, such that it can be easily contacted during the lithography process.

For double gated devices, the relative sizes of the graphite gates need to be considered. In order to contact the back gate in the lithography process, the bottom gate should also be longer than the top gate. On the other hand, the top gate should be wider than the back gate. This way the region that is gated only by the top gate can be also gated with the Si gate. This is very important in MATBG due to the existence of highly resistive states which can completely dominate the measured signatures otherwise. These considerations are displayed in Suppl. Fig. S3.

**Creating two graphene sheets with identical crystallographic orientation -** MATBG devices are always assembled starting from a single crystal graphene sheet, that is cut into two pieces. That ensures that both sheets have exactly the same initial crystallographic orientation prior to the rotation of the layers. The devices are fabricated using a cut-and-stack technique[22], where the original graphene flake is cut into two pieces. This approach has a big advantage to the original tear-and-stack method[28,38], as it does not induce a pulling and tearing motion in the graphene sheet during the pick-up process, and so reduces the chance of altering the relative twist angle between the layers. We use two different techniques to cut the graphene, one with an AFM cantilever, that is mounted on a glass slide, and the other with an ultra-strong pulsed laser beam.

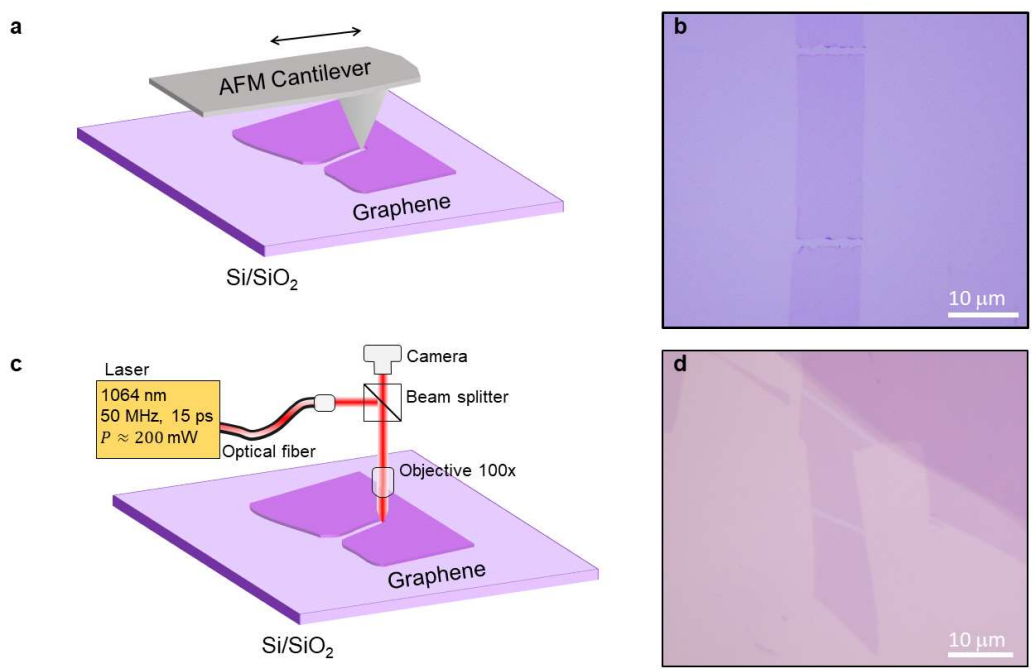

**Fig 3. AFM cantilever and laser cutting. a,** Schematic of the AFM cantilever on PDMS used to cut the graphene flake. **b,** The AFM cantilever is used to cut the graphene in-situ in the transfer stage. **c,** Schematic of the laser set-up used to cut the graphene flake. **d,** Graphene flake cut with the laser.

The first technique is built upon the typical stamping cantilevers for the stacking process: a small PDMS square is placed on a glass slide and an AFM cantilever is placed on the edge of the PDMS and secured with scotch tape (Fig. 3a). To cut the graphene, the glass slide with the AFM cantilever is placed on the micromanipulator of the transfer stage and lowered towards the chip with the desired graphene flake until contact is made. The point of contact can be seen as the cantilever deflects, changing its reflectance. Once the cantilever is in contact with the $SiO_2$/Si chip close to the desired graphene flake, the sample stage is moved passing the cantilever over the flake, which results in a clean-cut ca. 1 μm wide (Fig. 3a-b.). Graphene is cut at room $T$ to avoid sudden relaxation of the graphene flake, as at higher $T$ it tends to fold onto itself.

In the second technique, the graphene flake is cut by using a laser[39]. We use an infrared pulsed laser (1064 nm) with average power of 200 mW. The laser passes through a beam splitter which allows to focus it on the sample while imaging through the camera. The laser path which goes to the sample is focused using a 100x objective, which creates a beam size of ~ 1 μm width. By passing the laser through the desired flake, we can acquire clean cuts of also ca. 1 μm width (Fig. 3c-d). While both techniques give similar results, in principle, the laser induces less mechanical stress than the AFM cantilever, reducing the chance of breaking a flake while cutting it, making the graphene-cutting process more controlled.

**Cleaning of the area that surrounds the flake -** The AFM cantilever can also be used to move flakes[40,41]. This is especially useful when a flake is very close to the desired flake, such that it might negatively affect the pick-up process. By precisely controlling the AFM cantilever with the micromanipulators, one can fully push out a flake from the area as shown in Suppl. Fig. S4. Having a cleaner surface around the desired flakes helps to ensure a slow and controlled lamination of PC over the flakes (see Suppl. Video 1), which prevents the appearance of bubbles, helps squeeze any present bubbles out and lowers the chance of distorting the aimed twist-angle. Using the laser as described above, large graphene areas can be "burned", which can be useful to isolate a flake for pick up (shown in Suppl. Fig. S5). However, the laser cannot be used so far to structure or remove hBN flakes due to their chemical and temperature stability.

## 3. High-yield assembly of MATBG devices employing twist-angle locking

From the prior prepared catalogue of available 2D crystals we carefully select the best fitting flakes and prepare a tentative assembly plan of the ultimate stack. This allows to properly choose the size, shape and compatibility of different flakes to be carefully considered, and minimizes the possible errors arising during the stacking process.

**Preparation of the stacking process -** After cutting the graphene, pre-selecting all the flakes and making a stacking plan, the stacking process may begin. For the dry-transfer process we use a so-called stamp that is mounted on a glass slide. The stamp is a polymer heterostructure consisting of a small square of ca. 2 x 2 mm$^2$ of 1 mm thick commercially available polydimethylsiloxane (PDMS) that is covered with a polycarbonate (PC) film. The stamp is made following the work of Zomer et al.[27]. The full details of the stamp-making process are explained in Suppl. Section D.

The PDMS acts as a soft viscoelastic cushion in the pick-up process and the small size (2 x 2 mm$^2$) is chosen such that the contact point of the PC with the $SiO_2$ surface, which we generally refer to as the wavefront, can be controlled easily. The decision to use PC films as the adhesive layer is motivated mainly by its high adhesion properties to the used materials and because it permits to perform the stacking process at higher $T$ than f.e. polypropylene carbonate (PPC) thin films [42]. As is explained in detail later, the higher temperatures during the lamination process enhances the quality of the resulting device.

**Pick-up technique -** The first step in the assembly process of the 2D crystals, is to locate a clean region on the PC film, which is larger than the largest 2D crystal that will be used in the entire stack. Once this is chosen, the top graphite gate is picked up. The direction with which the PC is approaching the crystal in every step is important since it marks the relative orientation between the flakes. Hence, for the pick-up of the different crystals, we always rotate it into the ideal position.

The approach to pick up all the flakes, is as follows: The chip with the desired flake is placed on the heated sample stage and kept in place by applying vacuum on its back side. The stamp is lowered until contact is made with the heated $SiO_2$ surface. The point of contact is evident from the change in the deeper apparent color of the contact area, which is surrounded by Newton's rings (see Fig. 4a, d and g). In general, we set the tilt angle of the glass slide such that the PDMS/PC stamp makes its first contact with the $SiO_2$ in one of its corners, which allows for a better control of the wavefront. A sudden "jump", or fast movement of the PC film, can tear, move or induce bubbles in the heterostructure. Once the PC film has fully laminated over the flake, the stamp is pushed slightly further and then retracted slowly. When the flake is picked up, the PDMS/PC film will acquire a dark shadow in the shape of the flake, unlike the characteristic color it has on the $SiO_2$ surface.

**Locking the twist-angle of MATBG by anchoring to the hBN edges –** One issue that is commonly observed in the assembly of MATBG devices, is that the individual crystals are moving and rotating with respect to one another during the pick-up process. This has its roots in the lateral and vertical forces that are applied on the 2D crystals during the process, which can lead to a relative motion between the crystals that is enabled by the slippery and low friction van der Waals interfaces between them[43]. Especially between hBN, graphene and graphite flakes, this motion can lead to a distortion of the target twist-angles and positions of the 2D crystals in the stack. Furthermore, two graphene sheets are only energetically stable in the AB stacking configuration of $\theta = 0$ degrees twist-angle, where twisted bilayer graphene devices with $\theta \neq 0$ exist only in an energetically metastable state, and tend to rotate back to an AB configuration. These properties of vdW interfaces significantly lower the yield of a precise setting to the desired twist-angle between two graphene sheets.

In order to increase the yield of MATBG devices, it is therefore essential to develop a technique that hinders the free relative motion of the 2D crystals during the pick-up process and mechanically stabilizes them. For this purpose, we make use of a vdW edge clamping technique, which allows to interlock the edges of the individual layers, hence locking the relative twist-angle between them. As discussed prior, the first picked-up hBN is generally chosen to have at least one sharp edge, which will be used as an anchor, to which we clamp the edges of the two graphene flakes, which were defined in the graphene cutting process. The edge between the 2D crystals fold over each over and interlock over a length of ~ 1 μm, which is visible in the optical images (see Fig. 4b-d and Suppl. Fig. S10), and so restrict any further relative motion between them. This significantly increases the probability to retain a twist-angle of the TBG stack close to the magic-angle.

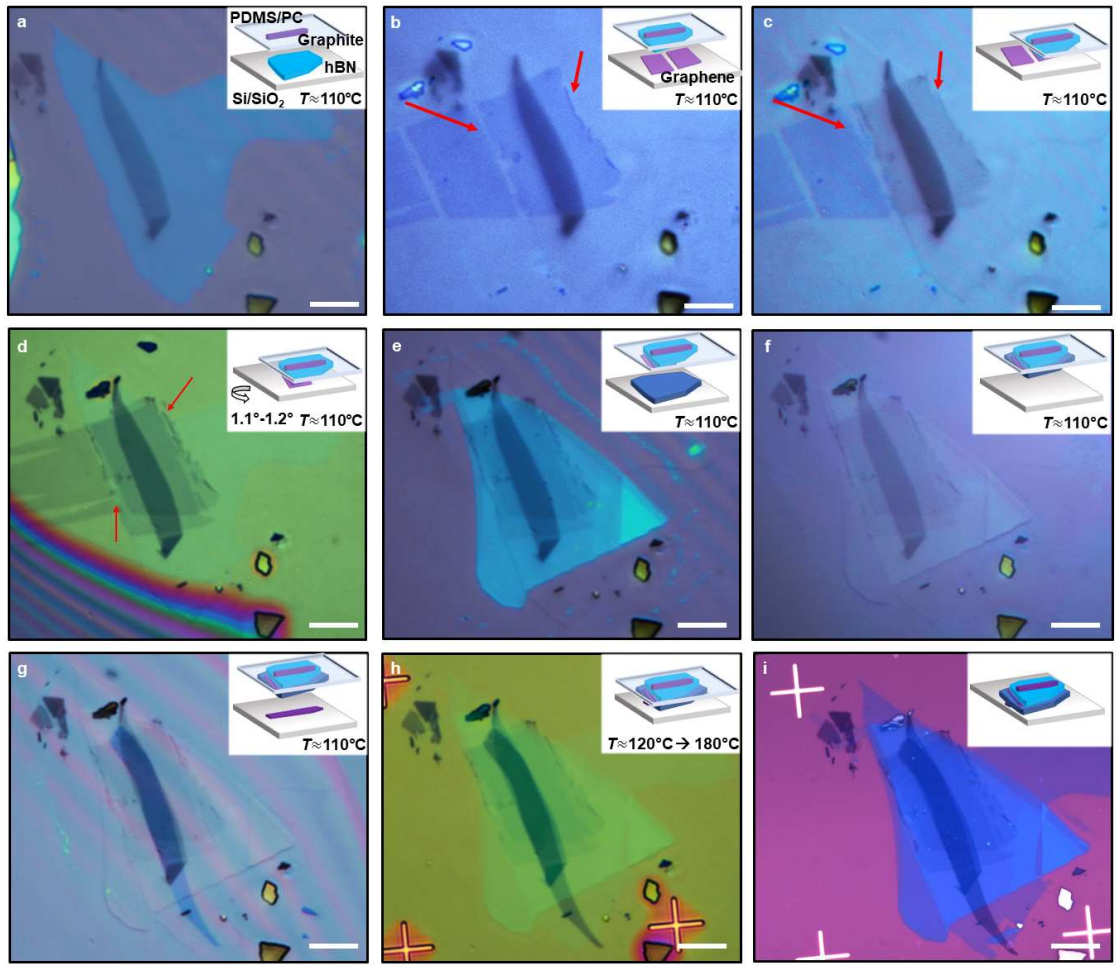

**Fig. 4. Stacking process. a,** Picking up the top hBN, while the top graphite is already picked up. **b,** Aligning the top hBN with the 1st graphene flake. The arrows signal the part of the graphene that will clamp over the hBN. **c,** 1st graphene is picked up. The change of color signals that the pick-up was successful. The red arrow points to the edges having "clamped" over the hBN. **d,** The stamp is laminated over the 2nd graphene after rotating the stage 1.1°. Red arrows point to the places where the 2nd graphene will clamp over the 1$^{st}$ graphene. **e,** Both graphene layers are now picked up. **f,** Picking up the bottom hBN. **g,** Aligning the stack over the bottom graphite gate. **h,** The stack is dropped on a pre-patterned chip with markers. **i,** Clean stack after removing the PC. The final stack has an angle of 1.06° ± 0.02°. The scale bar in all figures is 20 μm.

Importantly, the clamping has to be done using a non-crystallographic axis of the graphene and/or hBN to avoid unintentional alignment between the layers, which would induce an additional moiré pattern and influence the electronic properties of the stack[5,44]. Therefore, the clamping is not done between perfectly straight edges (which point to possible crystallographic axes) but rather asymmetric edges of similar size. Using the cut edge of the first graphene is an ideal clamping point because the rippled graphene provides more roughness (as is clearly seen in Fig. 3b). Hence, one strategy is to cut the graphene not just in two pieces, but rather in three, giving a cut edge also for the second graphene layer, as is seen in Fig. 4b-d.

**Stacking process -** A complete stacking process is shown in Fig. 4, where we follow all the steps of the fabrication of a double gated MATBG device. The same stacking procedure can be however extended to the fabrication of twisted graphene multilayers, and TMD bilayers. The entire pick-up process is done by fixing the stage temperature to $T \sim 100 - 120$ °C. The lamination on the flakes is done at constant $T$, and the approaching of the crystals is done entirely by hand using the $z$-micromanipulator on the transfer stage (see Suppl. Section E for a full description of the transfer stage). We do not approach the crystals by ramping up the temperature of the stage, as is used by other recipes elsewhere[27]. After the graphite top gate and the top hBN layer are picked up with the above recipe, we can now continue and pick up the two graphene layers.

When picking-up the graphene, the chip is arranged such that the cut in the graphene is matched with the sharp edge of the top hBN (Fig. 4b), and the wavefront is parallel to the cut. This facilitates to fully laminate over the 1st graphene flake while avoiding any contact between the PC or the hBN with the 2nd graphene flake. The wavefront is approached very slowly to avoid any unintended movement of the graphene, as any movement of either the top or bottom graphene sheet can cause a distortion in the twist angle. Once the hBN is in contact with the graphene, the wavefront is further moved until the entire first graphene sheet is covered with hBN, while ensuring that the PC does not touch the second graphene. As soon as the first graphene sheet is in full contact, the stamp is slowly retrieved and moved a few mm above the Si chip. During this pick-up step, the graphene flake is clamped with the top hBN layer.

Now the second graphene sheet is picked up. While the stamp and the top half of the stack are hovering over the chip, the sample stage is rotated by 1.1 - 1.2° to a slightly higher angle than the target twist-angle, to account for an often observed slight twist-angle relaxation of $\sim 0.1°$ during the pick-up process. After rotating, the second layer of graphene is overlapped with the first graphene layer and the pick-up procedure is repeated in the same fashion, as is described above for the first layer of graphene. The second graphene layer is also clamped to the hBN edge, in exactly the same way as the first layer. After picking up both graphene layers, the bottom hBN is picked up. The flake is approached in such a way that it fully encapsulates the graphene and that it will fully cover the bottom graphite gate.

Finally, the bottom gate is picked up. The bottom graphite gate should be entirely covered by hBN. If not entirely covered, it may have different adhesive behaviors between the hBN-covered region and the region in direct contact with the polymer, consequently inducing tension or strain during the pick-up process. This tension can be so violent that it can sometimes even displace the position of the graphite gate, destroying the whole stack. We often have observed that this type of tension has affected the twisted bilayer graphene region, and has enabled the relaxation of its twist angle. In double gated devices this pick-up step is even more crucial since both of the gates need to be perfectly aligned in order to have a working device.

**Dropping the stack -** Finally, the complete stack is dropped on a $SiO_2$/Si chip with preformed alignment markers to facilitate the subsequent nanofabrication process. Before the drop, the chips are cleaned with $O_2$ plasma to improve the adhesion of the 2D layers. The contact between the PC film and the chip is now made at $T \sim 120-150$ °C to enhance bubble mobility one more time[32]. The wavefront is moved very slowly over the stack to push away all the remaining bubbles. Once the full stack is in contact with the $SiO_2$, the wave front is moved ca. 200 μm further from the stack. Now the stage temperature $T$ is raised slowly up to 180 °C. As the $T$ approaches the glass transition temperature, $T_g$, of the PC of $\sim 147°C$ [42] the PC detaches from the PDMS film and at a $T$ far beyond the glass transition $\sim 180$ °C, the PC completely melts. For our PC/PDMS stamps and transfer setup, the detaching happens at slightly lower $T$, typically at a setpoint of ~ 130 °C. At this point the $z$-micromanipulator is moved up slightly to detach the entire PC film from the PDMS. During this process (130 °C < $T$ < 180 °C), we make sure that the PDMS is not in contact with the PC film, and we move it slightly up every time it contacts it. Once $T$ reaches ~180 °C, the stamp is fully retracted. At this point, the areas of the PC film that are in contact with the chip are fully molten and detach from the remaining PC areas on the glass slide. A full example of the procedure is shown in Suppl. Video 2.

The $T$ ranges in this step are very important. Retracting too far at low $T$ can break the stack, while if $T$ is raised without detaching the PC from the PDMS, the thermal expansion of the latter can put pressure on the stack and thus relax the twist angle. During the entire process, the X-Y micromanipulator of the stamp and the sample stage should not be moved, since this will tear the stack. Once the stack is dropped, the $T$ of the stage is lowered to room temperature. The stacking process is now finished. The final step prior to the lithography is to clean the PC. The chip is dipped in chloroform for 2 min, followed by dipping in acetone for 1 min, isopropanol for 1 min, and blow drying with $N_2$.

**Etching and contacting -** After the preparation of the stacks, these are fabricated into a Hall bar geometry (as schematically shown in Fig. 1a) via nanolithography techniques, namely e-beam lithography, reactive ion etching and evaporation. The heterostructures are etched using $CHF_3/O_2$ plasma and the 1-D contacts are made using 5 nm Cr / 50 nm Au following the recipe of Wang et al.[26].

## 4. Strategies to enhance twist-angle homogeneity by reducing bubble formation

In order to achieve the cleanest 2D interfaces and highest twist-angle homogeneity possible, we aim to avoid bubbles from forming during the stacking process as much as possible, as has been explained above. Bubbles significantly contribute to angle inhomogeneity and can even lead to the absence of the magic-angle condition in an area up to ca. 0.5 μm around the bubble[21], and can induce quite strong strain field in the device. Typically, bubbles form during the stacking process mainly because of accumulation of dirt on the surfaces of the different 2D materials [32,45] or due to fast wave-front approaches, which can trap air along the interface[46]. The overarching theme is to maintain a stacking process that is as smooth as possible, which involves full control of the stamp's wavefront. Here we summarize the main strategies we use to reduce bubble formation:

**Using clean stamps and flakes** - We make sure we use an area of the stamp which is clean and larger than the largest flake to be used. If the stacking process is done in an area of the PC film which already has some dust particles, bubbles, etc. this could hinder the pick-up process and introduce bubbles into the whole stack. In the same manner, we only use flakes which have clean interfaces, as any dirt or defect present on the flakes will induce bubbles.

**Using thin hBN flakes** - We use thin hBN (10 – 20 nm), which is considerably more elastic than thicker flakes, and makes the stacking process much smoother and may generally help in the strain relaxation within the hetero-structure. It also aids to avoid unexpected rapid movements or "jumps" in the stamp during the stacking, which can give to sudden stress release and so enhance bubble formation [33] (an example is shown in Suppl. Video 1).

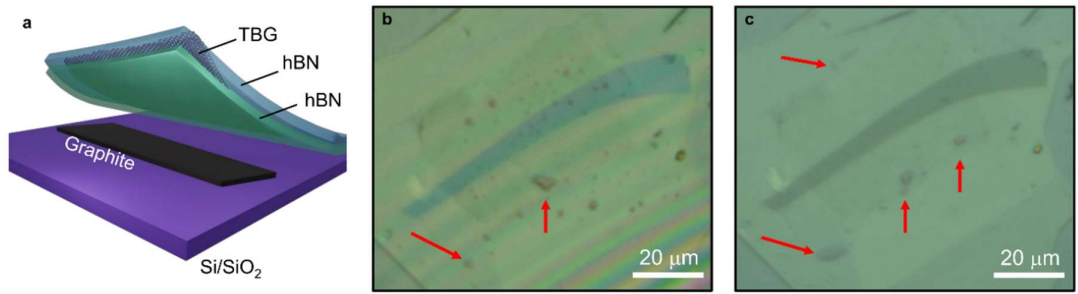

**Fig. 5. Bubble removal during the pick-up process. a,** Schematic showing the stacking process during the back gate pick-up step. **b,** Prior to picking up the back gate there are several bubbles visible inside the graphite gate area, which will be the device region. **c,** After picking up the back gate the bubbles are efficiently moved out of the device area and they accumulate at the edges of the graphene. The final device had a twist angle of $1.1° \pm 0.02°$.

**Cleaning surrounding areas** - We clean the area surrounding the desired flake (see Suppl. Section C). If there are large flakes or dirt close to the flake to be picked up, this modifies the wavefront and can easily lead to "jumps" in the stacking process.

**High temperature colamination process** - The entire stacking process is done at high $T \sim 100 - 120$ °C as this serves to improve the self-cleaning properties of the van der Waals interfaces during the stacking process and to enhance bubble mobility in all the pick-up steps[32,42]. This enables us to use every step to remove bubbles that might have formed in the previous pick-up steps. To highlight the bubble cleaning during different steps, we show the bubble removal in a single graphite gated device in Fig. 5. In the image we observe a stack which has many bubbles before the gate pickup, and during the pickup the bubbles rearrange, leaving the graphite region clean with fewer, larger bubbles being formed on the edge of the graphene.

### 5. Characterization of the devices

In order to understand the device homogeneity of our fabrication protocol, we have analyzed 34 so fabricated and finalized devices. First we studied their bubble density by optical microscopy, atomic force microscopy (AFM), and in a few devices, by scanning transmission electron microscopy (STEM).

**Optical microscopy** - By looking at 100x optical images we can count the amount of bubbles present in the stacks. We have compared some of the images to AFM scans of the same stack and found that with the right microscope settings we observe the same bubbles as in the AFM images (see Suppl. Fig. S11). We characterize the cleanliness of the stacks by extracting three main quantities: largest area with no bubbles, number of bubbles larger than 1 $\mu m^2$ (microbubbles) and number of bubbles smaller than 1 $\mu m^2$ (nanobubbles). The final area that we select for a device is always centered around the graphite gate, which is the region of interest to create a device. We have found that we obtain in average less than 1 microbubble and less than 3 nanobubbles per 10 x 5 $\mu m^2$, with several devices having no bubbles at all in that region and some devices reaching bubble free areas larger than 200 $\mu m^2$ (see Suppl. Fig. S12). Considering that the final devices are Hall bars with dimensions ranging 8 – 15 $\mu m$ x 2 - 4 $\mu m$, this procedure allows us to make the final devices in an entirely bubble free region of the stack.

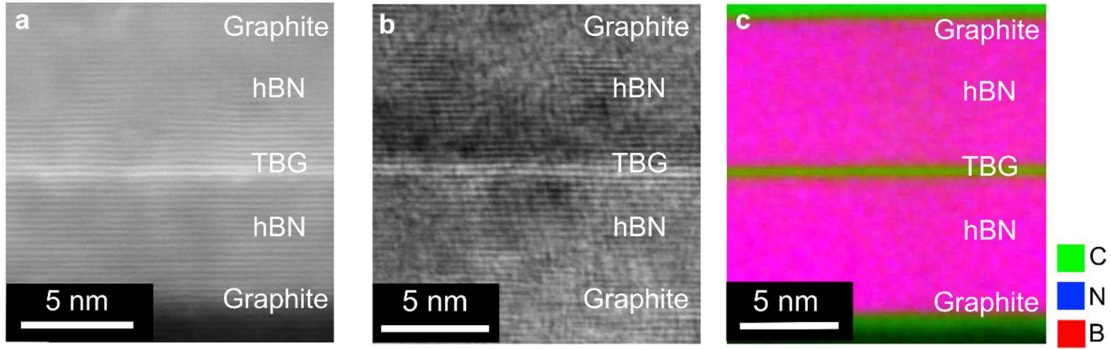

**Fig 6. Cross sectional STEM and EELS imaging of remote device regions. a-b,** STEM crossectional images of the device, taken in different regions of the device. The lattice fringes of the double-layer graphene sandwiched between the hBN layers, as well the graphite electrodes, are clearly visible in the STEM images. **c,** STEM-EELS mapping of the device cross section showing the graphite, hBN and double-layer graphene stacking in the device. In most of the device areas, the graphene/graphene and graphene/hBN, as well as hBN/graphite, interfaces are homogeneous and clean at the atomic scale.

**STEM microscopy** - The bubble density found by optical microscopy is further confirmed by scanning transmission electron microscopy (STEM). In Fig. 6 we show several STEM images taken along the longitudinal cross section of a device. By combining the STEM images with an electron energy loss spectroscopy (EELS) composition mapping, we can clearly see the different layers conforming the stack, like the hBN, graphite and MATBG sheets. For one of the devices we find no bubbles across the length of the entire device of ~12 μm, while for the other device (shown in Suppl. Fig. S13) we observe most of the device to be bubble free over a length of ~15 μm, but in some remote regions still find a few bubbles of ~ 50 nm in size.

**Room temperature transport characterization** – To select the most promising devices for further studies, we perform 4-terminal resistance ($R_{xx}$) measurements as a function of back gate voltage $V_g$ at room temperature. Due to the strong dependence of the TBG band-structure on the twist-angle, the characteristic gate sweeps of $R_{xx}$ vs. $V_g$ allow to distinguish between devices with a twist-angle close to the magic-angle, from devices with a lower ($\theta \lessapprox 0.7°$), higher ($\theta \gtrapprox 1.6°$) and completely relaxed twist-angle of ($\theta \approx 0°$). At the magic angle condition, the bands at the Fermi energy are highly non-dispersive (flat-bands) and are separated by the dispersive bands by a gap of ~ 40 meV. As the angle increases, the bands at the Fermi level become more dispersive while the band gaps move to higher energies[4,47]. On the other hand, for low twist-angles, several flat bands appear at low energies with small band gaps between them[48–50].

The room $T$ measurements can clearly resolve these signatures which distinguish between devices, by two main features: the shape of the $R_{xx}$ vs. $V_g$ dependence and the nominal value of the $R_{xx}$ (shown in Fig. 7). While devices with $\theta \approx 0°$ have a very sharp charge neutrality point (CNP), a similar but broadened behavior is observed for TBG with $\theta \lessapprox 0.7°$ due to the presence of multiple bands close to the Fermi level. As the angle gets larger ($\theta \gtrapprox 1.6°$), the band gap to the dispersive bands move to higher energies, leading to a characteristic double-humped curve (see Suppl. Fig. S14 for an example). Devices with $\theta$ close to the magic-angle have a much broader, dome shaped, $R_{xx}$ vs. $V_g$ dependence and a characteristically high $R_{xx} \gtrapprox 10$ kΩ.

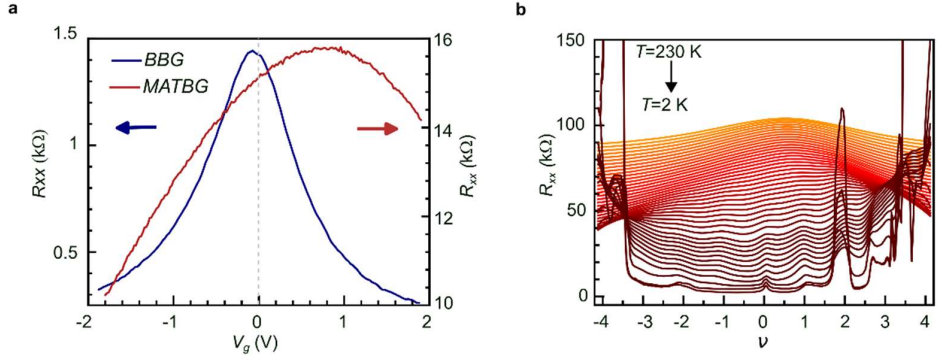

**Fig. 7. Identifying MATBG devices at room $T$. a,** Room $T$ measurement comparing a Bernal bilayer graphene (BBG) and a magic-angle twisted bilayer graphene (MATBG) device with twist angle $\theta = 1.06 \pm 0.02°$, shown in the blue and red curves, respectively. The asymmetry of the dome, combined with the 4-terminal resistance ($R_{xx}$) value allows to distinguish between the two. hBN of comparable thicknesses ($d \approx 15$ nm) are used in both devices. **b,** Cooldown curves of the MATBG shown in **a** from 230 K down to 2 K. The curves are shifted by 1 k$\Omega$ starting from the lowest $T$ for clarity.

One can attribute this behaviour to the simultaneous conduction through a flat-band and two temperature activated dispersive bands. However, we also find that devices with an intermediate large/small twist angle (1.3° ⪅ $\theta$ ⪅ 1.6° and 0.7° ⪅ $\theta$ ⪅ 0.9°) are nearly indistinguishable from magic-angle devices at room $T$.

**Low temperature transport characterization** – The devices are then cooled down, where typically below a $T < 100$ K the $R_{xx}$ vs. $\nu$ curves start to alter significantly from its high temperature shape, and below a $T < 10$ K the $R_{xx}$ vs. $\nu$ curves are dominated by the formation of the insulating and superconducting states, as can be seen in Fig. 7 b. These states are further characterized at low temperatures, typically at a base temperature of $T = 35$ mK, as has been shown in Fig. 1 b and c.

While it unfortunately is not possible to directly image the moiré pattern in a fully encapsulated and top-gated device, it is however possible to infer the twist-angle from the magneto-transport $R_{xx}$ vs. $B$ and $\nu$ phase-diagram as previously described[1,2] (see the Methods section for details). By measuring the carrier concentration $n$ in the device, through Hall and Shubnikov-de Haas experiments, it is possible to extract the size of the moiré unit cell, the corresponding filling factor $\nu$, and so assign the average twist-angle $\theta$ between the used contact pairs in the measurement. We then can perform two terminal conductance measurements between all the different contact pairs, and use the carrier concentration at which the correlated insulators at $\nu = 2$ appear, in order to estimate the average twist-angle between all the different regions in the device, as is shown in Fig. 8.

## 6. Device yield and twist-angle homogeneity

Finally, we summarize the main results that we can extract from 34 devices fabricated with the provided protocol, mainly the success rate or yield of the MATBG devices that have a twist-angle close to the magic-angle of $\theta = 1.1°$, and their twist-angle homogeneity $\Delta\theta$. For the twist angle homogeneity, we typically measure the 2-probe terminal conductance between all the contact pairs (generally separated by 1.5 - 2 μm), and extract the twist-angle $\theta$ using the carrier density $n$ at which the CI state at $\upsilon = 2$ nucleates, which is marked by a sharp dip in the conductance $G$. We have then estimated the number of contact pairs with a variability of twist angle $\Delta\theta \leq 0.02°$, which is the typical error bar of the angle extraction process as explained in the Methods section.

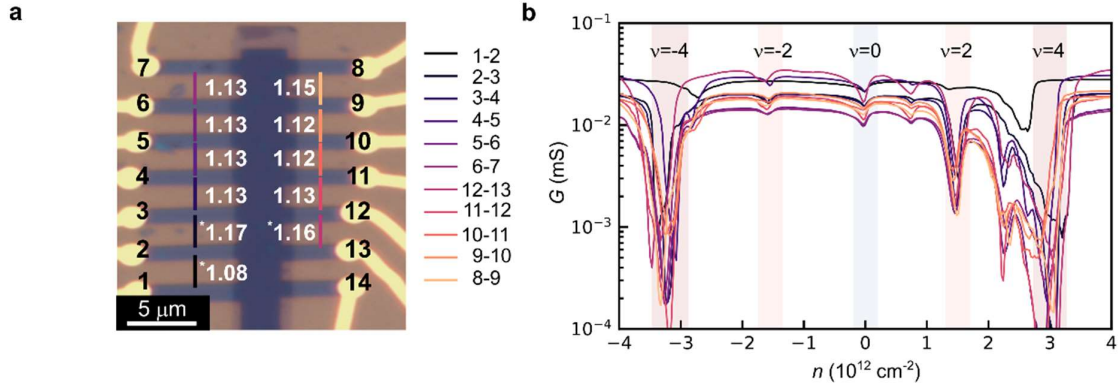

**Fig. 8. Low-temperature transport extraction of the twist-angles. a,** Optical image of a MATBG device with twist-angle homogeneity over a range of 36 μm². **b,** Two probe terminal conductance $G$ vs. carrier density $n$ measurement at $T = 35$ mK for the different contacts in the device shown in **a**.

**MATBG device yield** – To extract the MATBG device yield we have considered only samples that were made after the establishing of this fabrication protocol, that were fabricated by 3 fully-trained and experienced PhD students (the first 3 authors in this paper) in the period from 2020 – 2023. This accounted for a total of ~ 56 attempted stacks, of which 34 were fully fabricated and measured. Out of the 34 finalized devices, 13 (38%) had at least one region between two contact pairs with a twist angle $\theta = 1.1 \pm 0.1°$, out of which 8 devices (62%) showed superconductivity and 11 devices (85%) showed a correlated insulator state at a filling either $\upsilon = +2$ or $\upsilon = -2$. Another 5 devices showed a twist-angle between $\theta = 1.1 \pm 0.2°$, totaling a device yield of (53%) for this range of twist-angles. Devices which failed at an earlier stage of the fabrication process (ca. 40%) were not counted for the yield calculation.

**Twist-angle homogeneity** – By extracting the twist-angles between all of the different contact pairs in the devices, we typically find regions of about 6 μm² that show almost no twist-angle variation, of only $\Delta\theta \leq 0.02°$, with some devices reaching homogenous areas of up to 36 μm², as is highlighted for a best-case device shown in Fig. 8. As also seen from the twist-angle distribution in this device, we also observe sudden jumps in the twist-angles of neighboring contact pairs, likely introduced by some fault lines between the 2 graphene sheets.

It is important to note that the twist-angle extracted in transport measurements is a global twist angle determined by the average carrier density ($\overline{n_e} \approx C_g V_g/e$) on the micrometer scale. While we observe a homogenous global twist angle, it is likely that in the nanometer scale (few moiré unit cells), there are areas with different twist angles $\theta(r)$, which translate into a local carrier density distribution $n(r)$[21,51]. While these local twist angle/carrier density inhomogeneities do not affect the global twist angle extraction done by the average carrier density of the correlated and band insulating states (see Methods), they can affect the more fragile states such as superconductivity or magnetism [52].

# 7. Conclusion

While the field of MATBG specifically and that of moiré materials more generally, offers immense new opportunities to uncover exotic quantum phases[53,54], the device fabrication remains tedious and prone to strong twist-angle disorder. The fabrication of homogeneous and highly reproducible devices remains a big challenge and the main limiting factor of the advancement of the field. Although efforts are being made towards improving the reliability in the fabrication, for example by the automatization of the stacking process[55,56] or developing assembly processes in high vacuum for improving the cleanliness[57], a deeper understanding towards the subtleties of the stacking process is needed to progress further in that direction. The detailed fabrication process for MATBG devices presented here, allows to further standardize the assembly of moiré materials by using an almost bubble-free assembly protocol and a twist-angle clamping technique, that allows to obtain large homogeneous device regions with the desired twist-angle and with a high probability. The very controlled lamination process, aided by the use of thin hBN, may also contribute to reduce the accumulated strain within the heterostructure. We are certain that this protocol can be used for all other moiré materials, and will increase their device yield and twist-angle homogeneity.

**References:**

1. Cao, Y. *et al.* Correlated insulator behaviour at half-filling in magic-angle graphene superlattices. *Nature* **556**, 80–84 (2018).
2. Lu, X. *et al.* Superconductors, Orbital Magnets, and Correlated States in Magic Angle Bilayer Graphene. *Nature* **574**, 653 (2019).
3. Yankowitz, M. *et al.* Tuning superconductivity in twisted bilayer graphene. *Science* **363**, 1059–1064 (2019).
4. Cao, Y. *et al.* Unconventional superconductivity in magic-angle graphene superlattices. *Nature* **556**, 43–50 (2018).
5. Sharpe, A. L. *et al.* Emergent ferromagnetism near three-quarters filling in twisted bilayer graphene. *Science* **365**, 605–608 (2019).
6. Serlin, M. *et al.* Intrinsic quantized anomalous Hall effect in a moiré heterostructure. *Science* **367**, 900–903 (2020).
7. Tseng, C.-C. *et al.* Anomalous Hall effect at half filling in twisted bilayer graphene. *Nat. Phys.* **18**, 1038–1042 (2022).
8. Nuckolls, K. P. *et al.* Strongly correlated Chern insulators in magic-angle twisted bilayer graphene. *Nature* **588**, 610–615 (2020).
9. Das, I. *et al.* Symmetry-broken Chern insulators and Rashba-like Landau-level crossings in magic-angle bilayer graphene. *Nat. Phys.* **17**, 710–714 (2021).
10. Wu, S., Zhang, Z., Watanabe, K., Taniguchi, T. & Andrei, E. Y. Chern insulators, van Hove singularities and topological flat bands in magic-angle twisted bilayer graphene. *Nat. Mater.* **20**, 488–494 (2021).
11. Saito, Y. *et al.* Hofstadter subband ferromagnetism and symmetry-broken Chern insulators in twisted bilayer graphene. *Nat. Phys.* **17**, 478–481 (2021).
12. Cao, Y. *et al.* Strange Metal in Magic-Angle Graphene with near Planckian Dissipation. *Phys. Rev. Lett.* **124**, 76801 (2020).
13. Polshyn, H. *et al.* Large linear-in-temperature resistivity in twisted bilayer graphene. *Nat. Phys.* **15**, 1011–1016 (2019).
14. Jaoui, A. *et al.* Quantum critical behaviour in magic-angle twisted bilayer graphene. *Nat. Phys.* *2022* 1–6 (2022) doi:10.1038/S41567-022-01556-5.
15. Wang, L. *et al.* Correlated electronic phases in twisted bilayer transition metal dichalcogenides. *Nat. Mater.* **19**, 861–866 (2020).
16. Li, T. *et al.* Continuous Mott transition in semiconductor moiré superlattices. *Nature* **597**, 350–354 (2021).
17. Khalaf, E., Kruchkov, A. J., Tarnopolsky, G. & Vishwanath, A. Magic angle hierarchy in twisted graphene multilayers. *Phys. Rev. B* **100**, 085109 (2019).
18. Zhang, Y. *et al.* Promotion of superconductivity in magic-angle graphene multilayers. *Science* **377**, 1538–1543 (2022).
19. Park, J. M. *et al.* Robust superconductivity in magic-angle multilayer graphene family. *Nat. Mater. 2022 218* **21**, 877–883 (2022).
20. Balents, L., Dean, C. R., Efetov, D. K. & Young, A. F. Superconductivity and strong correlations in moiré flat bands. *Nat. Phys.* **16**, 725–733 (2020).
21. Uri, A. *et al.* Mapping the twist-angle disorder and Landau levels in magic-angle graphene. *Nature* **581**, 47–52 (2020).
22. Stepanov, P. *et al.* Untying the insulating and superconducting orders in magic-angle graphene. *Nature* **583**, 375–378 (2020).
23. Arora, H. S. *et al.* Superconductivity in metallic twisted bilayer graphene stabilized by WSe2. *Nature* **583**, 379–384 (2020).
24. Choi, Y. *et al.* Electronic correlations in twisted bilayer graphene near the magic angle. *Nat. Phys.* **15**, 1174–1180 (2019).
25. Kazmierczak, N. P. *et al.* Strain fields in twisted bilayer graphene. *Nat. Mater.* **20**, 956–963 (2021).

26. Wang, L. *et al.* One-dimensional electrical contact to a two-dimensional material. *Science* **342**, 614–617 (2013).
27. Zomer, P. J., Guimarães, M. H. D., Brant, J. C., Tombros, N. & Van Wees, B. J. Fast pick up technique for high quality heterostructures of bilayer graphene and hexagonal boron nitride. *Appl. Phys. Lett.* **105**, (2014).
28. Kim, K. *et al.* Van der Waals Heterostructures with High Accuracy Rotational Alignment. *Nano Lett.* **16**, 1989–1995 (2016).
29. Saito, Y. *et al.* Isospin Pomeranchuk effect in twisted bilayer graphene. *Nat. 2021 5927853* **592**, 220–224 (2021).
30. Rozen, A. *et al.* Entropic evidence for a Pomeranchuk effect in magic-angle graphene. *Nature* **592**, 214–219 (2021).
31. Huang, Y. *et al.* Reliable Exfoliation of Large-Area High-Quality Flakes of Graphene and Other Two-Dimensional Materials. *ACS Nano* **9**, 10612–10620 (2015).
32. Pizzocchero, F. *et al.* The hot pick-up technique for batch assembly of van der Waals heterostructures. *Nat. Commun. 2016 71* **7**, 1–10 (2016).
33. Castellanos-Gomez, A. *et al.* Deterministic transfer of two-dimensional materials by all-dry viscoelastic stamping. *2D Mater.* **1**, 011002 (2014).
34. Britnell, L. *et al.* Electron tunneling through ultrathin boron nitride crystalline barriers. *Nano Lett.* **12**, 1707–1710 (2012).
35. Zhou, H. *et al.* Half- and quarter-metals in rhombohedral trilayer graphene. *Nature* **598**, 429–433 (2021).
36. Miyazaki, H. *et al.* Inter-Layer Screening Length to Electric Field in Thin Graphite Film. *Appl. Phys. Express* **1**, 034007 (2008).
37. Han, E. *et al.* Ultrasoft slip-mediated bending in few-layer graphene. *Nat. Mater.* **19**, 305–309 (2020).
38. Cao, Y. *et al.* Superlattice-Induced Insulating States and Valley-Protected Orbits in Twisted Bilayer Graphene. *Phys. Rev. Lett.* **117**, (2016).
39. Park, J. M., Cao, Y., Watanabe, K., Taniguchi, T. & Jarillo-Herrero, P. Flavour Hund's coupling, Chern gaps and charge diffusivity in moiré graphene. *Nature* **592**, 43–48 (2021).
40. Ribeiro-Palau, R. *et al.* Twistable electronics with dynamically rotatable heterostructures. *Science* **361**, 690–693 (2018).
41. Kapfer, M. *et al.* Programming twist angle and strain profiles in 2D materials. *Science* **381**, 677–681 (2023).
42. Purdie, D. G. *et al.* Cleaning interfaces in layered materials heterostructures. *Nat. Commun. 2018 91* **9**, 1–12 (2018).
43. Dai, Z., Lu, N., Liechti, K. M. & Huang, R. Mechanics at the interfaces of 2D materials: Challenges and opportunities. *Curr. Opin. Solid State Mater. Sci.* **24**, 100837 (2020).
44. Zhang, Y.-H., Mao, D. & Senthil, T. Twisted bilayer graphene aligned with hexagonal boron nitride: Anomalous Hall effect and a lattice model. *Phys. Rev. Res.* **1**, 033126 (2019).
45. Khestanova, E., Guinea, F., Fumagalli, L., Geim, A. K. & Grigorieva, I. V. Universal shape and pressure inside bubbles appearing in van der Waals heterostructures. *Nat. Commun.* **7**, 12587 (2016).
46. Haigh, S. J. *et al.* Cross-sectional imaging of individual layers and buried interfaces of graphene-based heterostructures and superlattices. *Nat. Mater.* **11**, 764–767 (2012).
47. Carr, S., Fang, S., Jarillo-Herrero, P. & Kaxiras, E. Pressure dependence of the magic twist angle in graphene superlattices. *Phys. Rev. B* **98**, 85144 (2018).
48. Bistritzer, R. & MacDonald, A. H. Moiré bands in twisted double-layer graphene. *Proc. Natl. Acad. Sci. U. S. A.* **108**, 12233–12237 (2011).
49. Lopes dos Santos, J. M. B., Peres, N. M. R. & Castro Neto, A. H. Continuum model of the twisted graphene bilayer. *Phys. Rev. B* **86**, 155449 (2012).

50. Lu, X. *et al.* Multiple flat bands and topological Hofstadter butterfly in twisted bilayer graphene close to the second magic angle. *Proc. Natl. Acad. Sci.* **118**, e2100006118 (2021).
51. Benschop, T. *et al.* Measuring local moir\'e lattice heterogeneity of twisted bilayer graphene. *Phys. Rev. Res.* **3**, 013153 (2021).
52. Grover, S. *et al.* Chern mosaic and Berry-curvature magnetism in magic-angle graphene. *Nat. Phys.* **18**, 885–892 (2022).
53. Carr, S. *et al.* Twistronics: Manipulating the electronic properties of two-dimensional layered structures through their twist angle. *Phys. Rev. B* **95**, 75420 (2017).
54. Geim, A. K. & Grigorieva, I. V. Van der Waals heterostructures. *Nature* **499**, 419–425 (2013).
55. Masubuchi, S. *et al.* Autonomous robotic searching and assembly of two-dimensional crystals to build van der Waals superlattices. doi:10.1038/s41467-018-03723-w.
56. Mannix, A. J. *et al.* Robotic four-dimensional pixel assembly of van der Waals solids. *Nat. Nanotechnol. 2022* 1–6 (2022) doi:10.1038/s41565-021-01061-5.
57. Wang, W. *et al.* Clean assembly of van der Waals heterostructures using silicon nitride membranes. *Nat. Electron.* **6**, 981–990 (2023).

## Methods:

<u>Transport measurements</u>: The room $T$ transport measurements were carried out in a home-made measurement set-up where the sample is placed under a vacuum ~$10^{-3}$ bar. Standard low-frequency lock-in techniques (Stanford Research SR860 amplifiers) were used to measure $R_{xx}$ with an excitation current of 10 nA at a frequency of 13.11 Hz. Keithley 2400 source-meters were used to control the gates. The low $T$ measurements were performed in a dilution refrigerator (Bluefors SD250) with a base temperature of 35 mK.

<u>Twist angle extraction</u>: The twist angle $\theta$ is extracted by applying the relation $n_s = 8\theta^2/\sqrt{3}a^2$, where $n_s$ is the superlattice carrier density and $a = 0.246$ nm is the graphene lattice constant. In order to accurately extract $n_s$, magnetotransport measurements, like the Landau-fan map ($R_{xx}$ vs. $V_g$ magnetic field $B$) shown in Fig. 1c, are used. First, the carrier density $n = C_g V_g/e$, is calibrated by extracting the capacitance $C_g$ from fitting the Landau levels arising from the CNP. Alternatively, the capacitance can be extracted using Hall measurements at low field. Near the CNP, the Hall carrier density $n_H = -B/eR_{xy}$, should closely follow the gate-induced carrier density $n_H = n$. Finally, the superlattice carrier density $n_s$ is extracted from the origin of the Landau levels emerging form the band insulators (BIs) or from the CI at half filling ($n_{\upsilon=2}$), such that $n_s=2n_{\upsilon=2}$. Since the twist angle extraction relies on the accuracy of the position of $n_s$, the calculated twist angles always have an error of ca. 0.02°.

<u>STEM imaging</u>: The STEM imaging and EELS measurements were performed using a JEOL monochromated ARM200F transmission electron microscope operated at 80 kV. The microscope is equipped with a Schottky field emission gun, a double-Wien monochromator, a CEOS ASCOR probe Cs corrector, a CEOS CETCOR image Cs corrector, and a Gatan image filter (GIF) Continuum for EELS, as well as ADF and bright field detectors for STEM imaging.

**Acknowledgements**

D.K.E. acknowledges funding from the European Research Council (ERC) under the European Union's Horizon 2020 research and innovation program (grant agreement No. 852927), the German Research Foundation (DFG) under the priority program SPP2244 (project No. 535146365), the EU EIC Pathfinder Grant "FLATS" (grant agreement No. 101099139) and the Keele Foundation. J.D.M. acknowledges support from the INPhINIT 'la Caixa' Foundation (ID 100010434) fellowship program (LCF/BQ/DI19/11730021). I. D. acknowledges funding from the German Research Foundation (DFG) under Germany's Excellence Strategy – EXC2111-390814868. K.W. and T.T. acknowledge support from the Elemental Strategy Initiative conducted by the MEXT, Japan (Grant Number JPMXP0112101001) and JSPS KAKENHI (Grant Numbers 19H05790, 20H00354 and 21H05233). L.J. Zeng and E. O acknowledge the financial support from the Knut and Alice Wallenberg Foundation (2019.0140). The authors acknowledge the financial support from Swedish Research Council (VR) and Swedish Foundation for Strategic Research (SSF) for the access to ARTEMI, the Swedish National Infrastructure in Advanced Electron Microscopy (2021-00171 and RIF21-0026).

# Supplementary information for "High-yield fabrication of bubble-free magic-angle twisted bilayer graphene devices with high twist-angle homogeneity"

J. Díez-Mérida[1,2,3], I. Das[2,3], G. Di Battista[2,3], A. Díez-Carlón[2,3], M. Lee[2,3], L. Zeng[4], K. Watanabe[5], T. Taniguchi[5], E. Olsson[4] and Dmitri K. Efetov[2,3]*

1. ICFO - Institut de Ciencies Fotoniques, The Barcelona Institute of Science and Technology, Castelldefels, Barcelona, 08860, Spain
2. Fakultät für Physik, Ludwig-Maximilians-Universität, Schellingstrasse 4, 80799 München, Germany
3. Munich Center for Quantum Science and Technology (MCQST), München, Germany
4. Department of Physics, Chalmers University of Technology, Gothenburg SE-41296, Sweden
5. National Institute of Material Sciences, 1-1 Namiki, Tsukuba, 305-0044, Japan

*E-mail: dmitri.efetov@physik.lmu.de

## Table of Contents

## A. Exfoliation process

**Graphene/graphite** - The procedure to exfoliate graphene/graphite is as follows:

- The Si/SiO$_2$ (285nm SiO$_2$) chips are cleaned with O$_2$ plasma for 3-5 min.
- A crystal of graphite is placed on a scotch tape. As the crystal is removed, it will leave a large piece of graphite on the tape.
- The tape is folded several times (~7-8 times) until most of it is covered in graphite flakes. The objective is to cover as much tape as possible with the minimum number of folds, as each fold will reduce the size of the potential graphene flakes.
- Once the tape is homogeneously covered in graphite flakes, it is pressed against the cleaned Si/SiO$_2$ chips to attach the flakes to the surface of the chips.
- The chips are then placed in a hot plate at ~105 °C for 2 min. The heating increases the contact between the flake and the SiO$_2$ by removing gas from the interface, and thus increasing the van der Waals forces between them. However, during heating the glue in the tape will also adhere to the SiO$_2$ surface, leaving unwanted residues around the graphene flakes. The longer the heating time the more residues there will be. In general, heating for 2- 3 min gives a good equilibrium between a high exfoliation yield and few residues, while heating for longer will give too many tape residues, therefore being detrimental for the process.
- After removing the tape from the hot plate, it is left to cool down for ~10-20 sec. Then the tape is peeled off from the chips very slowly. The slow motion is very important to avoid flakes from breaking, obtaining larger flakes. The waiting time before peeling the tape allows to remove the tape more slowly, as otherwise the glue will be too soft and the peeling off will be less controlled.

**hBN** - The hBN crystals are exfoliated in a very similar manner, with a few key differences:

- The original hBN crystals are much smaller than graphite, such that instead of starting with a large crystal, we start using several small hBN crystals.
- "Daughter" tape. The folding of the tape is done in a similar manner until the tape is heavily covered in smaller hBN crystals. However, in this case the crystals usually are too thick to directly exfoliate in the Si/SiO$_2$ chips. Therefore, in general a "daughter" tape is obtained by using a second tape and exfoliating from the first "mother" tape. The "mother" tape can be used for several "daughter" tapes. In general, if the crystals on the tape are very bright, this will lead to very thick hBN crystals which are not useful for stacking. Once the crystals in the tape have a more grayish, not-so-bright color, the crystals are too thin and they cannot be used anymore to create "daughter" tapes. Once the "daughter" tape is made, it is directly used to exfoliate on the Si/SiO$_2$ chips. Making the "daughter" tape denser by folding it again with itself is likely to break the crystals into very small pieces which are not useful for the stacking process.
- The chips are not heated prior to the exfoliation process. Although using a hot plate can increase the exfoliation yield, in the case of hBN the tape is not as dense as in the case of graphite, which means that there will be too many tape residues on the final chips. However, it is still important to wait a few minutes with the chips on the tape before removing the tape.
- When pushing the chips to the tape press gentler than in the case of graphene. if the hBN is pressed too hard the flakes will break leaving much smaller flakes than desired.

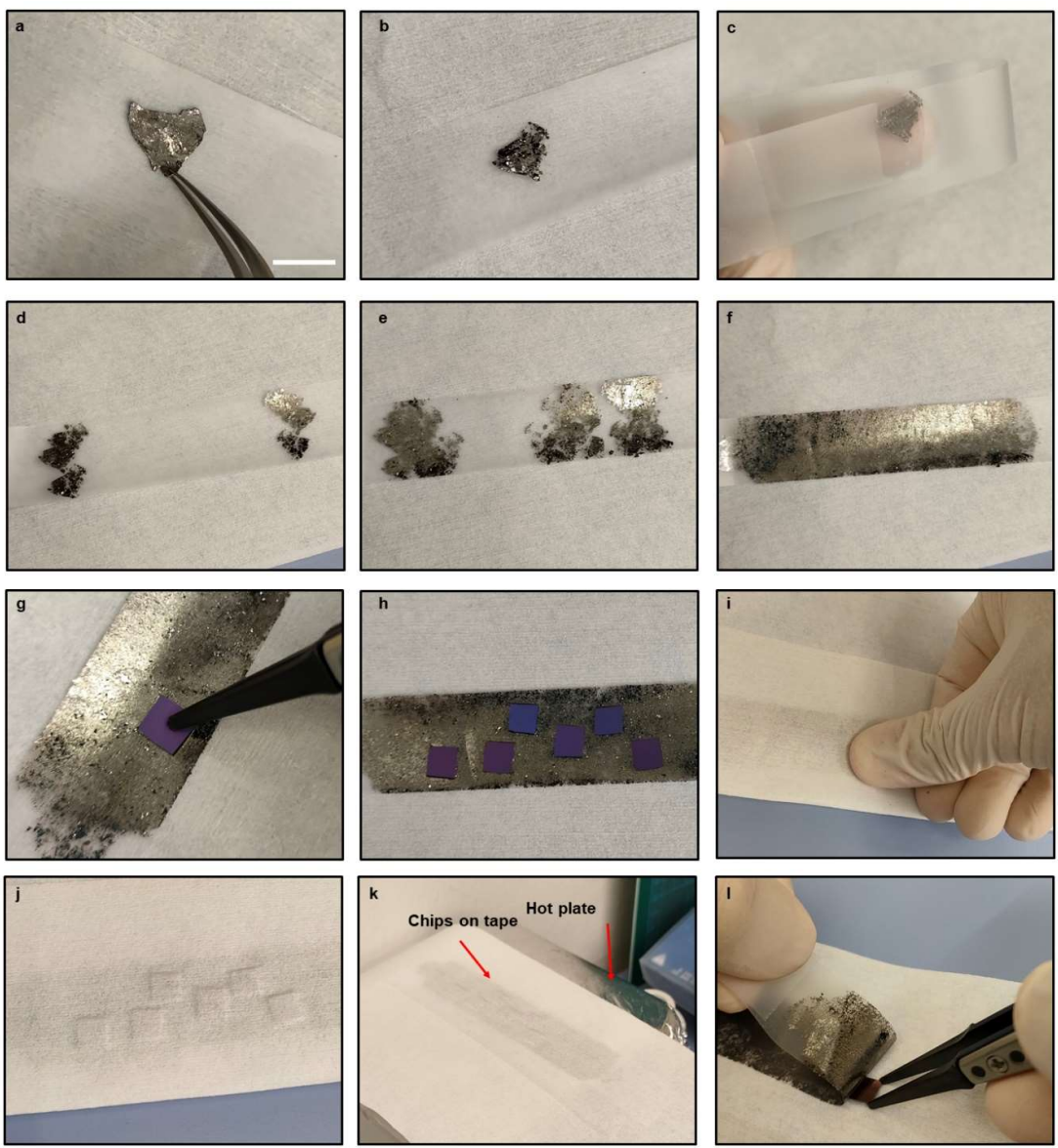

**Fig. S1. Graphene exfoliation. a,** A graphite crystal is placed on top of the scotch tape. **b,** The crystal is removed, leaving the exfoliated crystals on the tape. **c,** The tape is folded to quickly expand the crystal. Care is taken to touch only the area which has graphite to avoid extra residues. **d-f,** The crystals are exfoliated until covering the whole tape. **g,** A chip is placed on the tape region with graphite and gently pushed on the back. h, Several chips are placed on the tape. **i,** A soft tissue is used to press the chips on the graphite crystals to increase the exfoliation yield. **j,** After pressing the chips are properly attached to the tape with the graphite crystals. k, The chips are placed on a hot plate at ~105 °C for 2 min. **l,** Finally the chips are slowly peeled off from the tape. The scale bar in **a** is 5 mm.

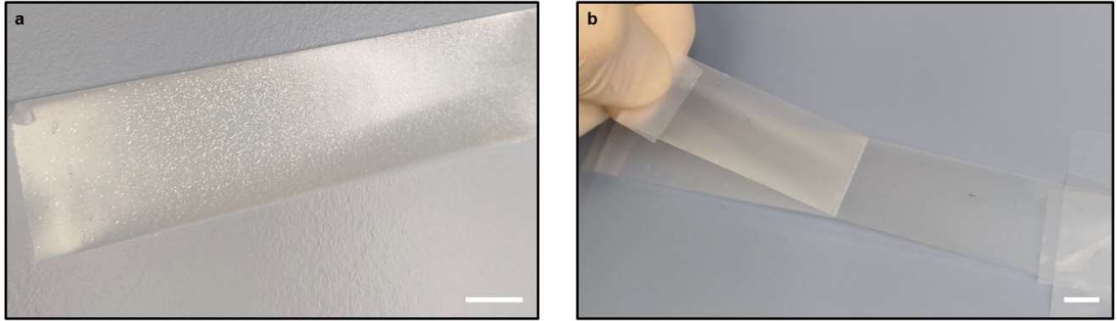

**Fig. S2. hBN exfoliation with "daughter" tape. a,** Original tape with the hBN exfoliated crystals. The tape is full oft crystals. **b,** A "daughter" tape is extracted from the original tape.

## B. Double graphite gate width ratios

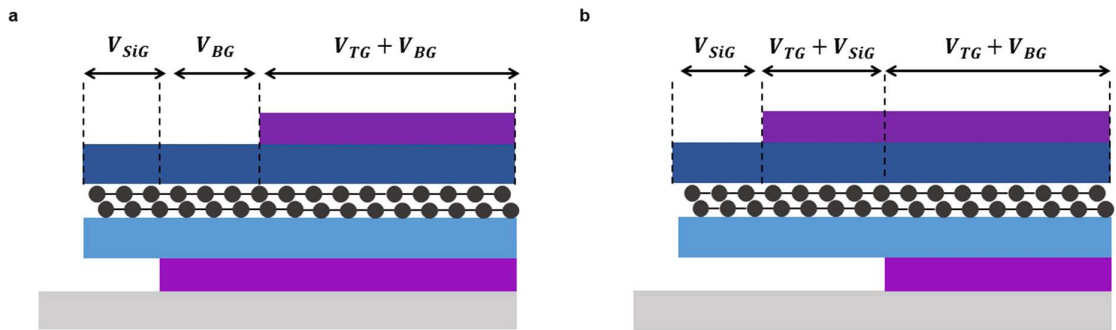

**Fig. S3. Effect of the width ratio between top and bottom gates**. $V_{SiG}$, $V_{BG}$ and $V_{TG}$ represent the regions gated by the silicon gate, the back gate and the top gate, respectively. **a,** Back gate is wider than top gate. There is a region where the back gate will screen the Si gate, creating extra features which cannot be removed. **b,** Top gate is wider than back gate. The extra region could now be removed by using the Si gate.

## C. Cleaning of the area that surrounds the flakes

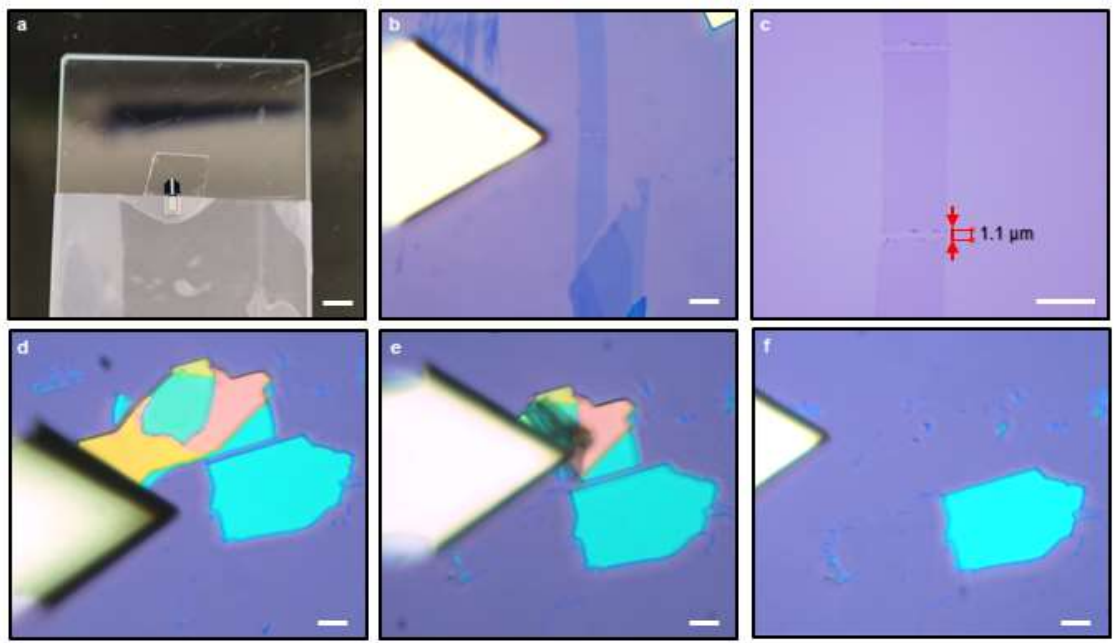

**Fig S4. AFM cantilever cutting and deterministic cleaning. a,** AFM cantilever on PDMS placed on a glass slide to be used in the transfer stage. **b,** The AFM cantilever is used to cut the graphene in-situ in the transfer stage. **c,** Zoom-in of the cut made in the flake in b. **d,** The AFM cantilever is used to remove the thicker flake beside the flake of interest. **e,** The second flake is being folded onto itself using the cantilever. **f,** The undesired flake is completely removed from the area. The good flake remains in place. The scale bar is 3 mm in **a** and 10 μm in the rest of the figures.

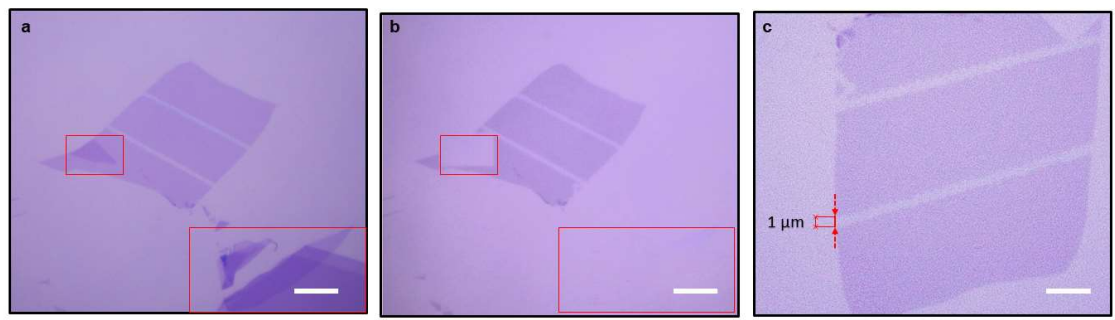

**Fig. S5. Laser cutting and burning of graphene. a,** Graphene flake which has been cut with an infrared (1064 nm) laser. **b,** The same laser can be used to clean away all the thicker graphite flakes around. This is shown in the bottom left corner of the graphene flake and also in the right bottom corner of the image (highlighted with the red squares). **c,** Zoom-in of the cuts, showing the width is approximately 1 μm. Scale bars are 10 μm in **a** and **b** and 5 μm in **c**.

## D. Stamp making

To make the PC film first a polycarbonate solution at 6 % weight in chloroform is made by introducing PC pellets in a beaker with chloroform and magnetically stirring overnight at room temperature. Once the PC is fully dissolved the solution is kept tightly closed and can be used for several weeks. The PC film is then made by transferring a few drops of the PC film into a glass slide. To have a homogenous film a second glass slide is pressed onto the first one and they are slid on top of each other leaving a homogenous film on both glass slides. We have found that films which give the best results are 2 – 3 μm thick. Finally, the glass slides with the PC film are put in a hot plate at ca. 100 °C for 2' to evaporate the excess chloroform and improve the homogeneity of the PC film. For the PDMS, commercially available PDMS films from Gelpak are used.

The procedure to make the actual stamps is shown in Fig. S2 and is as follows:
- A small square of PDMS of ~2 × 2 mm is placed on top of a clean glass slide.
- The PC film is cut into squares of ca. 1 ×1 cm.
- A hole larger than the PDMS square is made in a piece of scotch tape. The hole is used to expose the PC only in the region where there is PDMS below.
- A square of PC is picked up with the scotch tape and transferred on top of the PDMS. When transferring the PC film on top of the PDMS it should remain flat and relaxed, without visible wrinkles.
- The extra scotch tape is cut. The stamp is now finished.
- After finishing, the stamp is heated to 120 °C for ca. 5'. This will soften the PC, improving the conforming to PDMS and its adhesion to it.

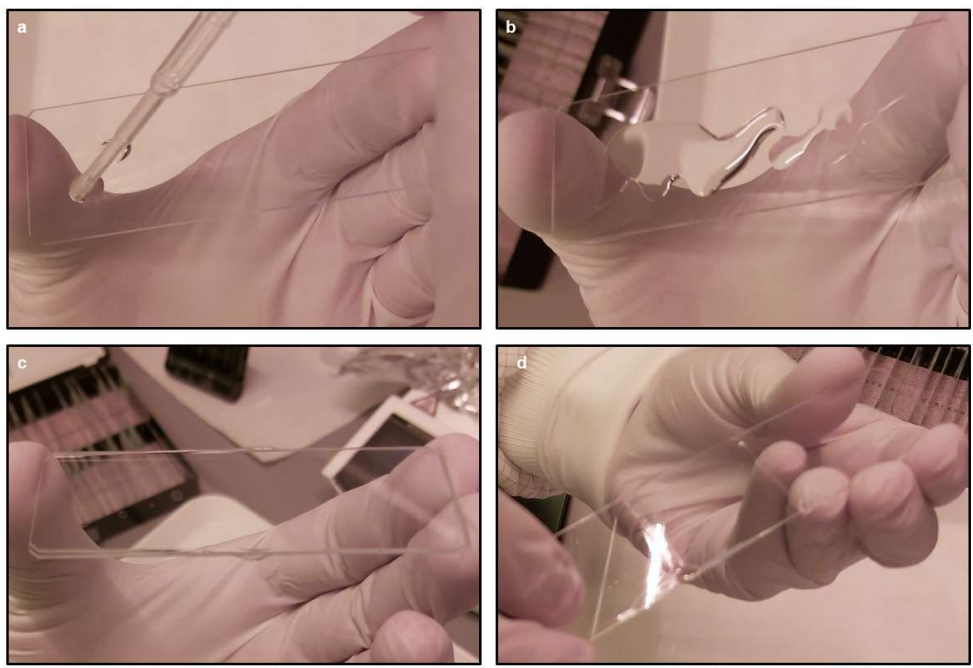

**Fig. S6. PC film making process. a,** A syringe is used to pour a few drops of PC solution on a glass slide. **b,** The PC solution poured on the glass slide. **c,** A second glass slide is placed on top of the one with the PC solution. **d,** The second glass slide is slowly released leaving a homogenous PC film on both surfaces.

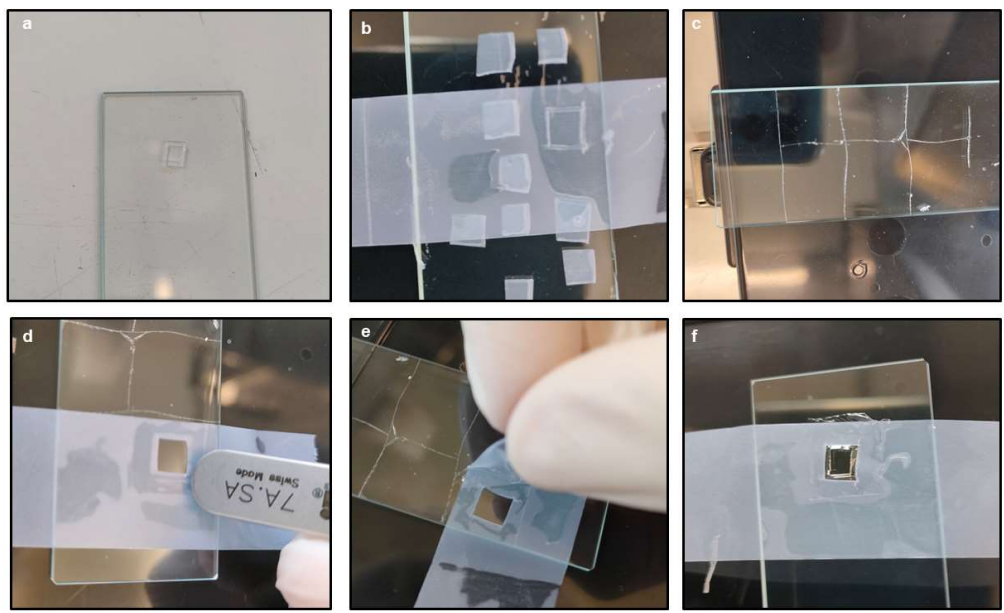

**Fig. S7. Stamp making process**. **a,** A PDMS square is placed on top of a clean glass slide. **b,** A hole larger than the PDMS square is cut on a scotch tape piece. **c,** The PC film is cut into small pieces to be picked up. **d,** The cut scotch tape is used to pick up the PC film. The back of the tweezers is used to ensure good contact between the tape and the PC film. **e,** The tape with the PC film is slowly released to avoid breaking it. **f,** The PC film is placed on top of the PDMS. The stamp is now finished.

### E. Vertical assembly setup

The dry transfer stacking technique is done using a so-called transfer stage or stamping setup. The transfer stage consists of a modified microscope in which the van der Waals structure assembly can be performed. The main components of the transfer stage are (see Fig. S5 for reference of the different parts):

- Heavy table for vibration stability and screw holes to hold the different pieces.
- Sample stage platform with X-Y and rotation control. In order to accurately rotate the two graphene sheets by an angle of 1.1°, the sample stage (b1) is placed on top of a goniometer (b2). The goniometer used has an angle control accuracy of 0.016°, giving a very precise angle of the rotation.
- Micromanipulator stage to control the stamps. A metallic "arm" (c4) which extends towards the sample stage where a glass slide with the stamp can be held during the assembly process. The X-Y manipulators (c1) are used to move the stamp around the sample to choose the right region and the Z manipulator (c2) is used to control the height and therefore make contact or retract the contact from the sample. An important feature of this stage is to have control over the tilt angle on the X-Y plane (c3). By setting the right angle is possible to control the angle at which the stamps will make the contact with the sample. This allows to control the stacking direction and the smoothness of the contact between the stamp and the sample. By having a very large tilt angle a large force will be put on the PC film, while having a low angle will not allow to control the point of contact, nor the wavefront of the PC film. The control over the wavefront is highly important since it determines the strain put on the flakes and also the smoothness of the PC approach to the flakes.
- Vacuum pump and valves. Vacuum is used to keep the sample, the stamp arm and the stamp in place during the stacking procedure. In the path between the pump and the final vacuum tubes in the stage, the tubes are reduced in diameter three times and the valves are held in a box. These processes completely isolate the vibrations from the pump. Care has to be taken to also isolate the electrical connections from the vibrations of the pump.
- Long working distance objectives. The microscope used for the transfer stage needs to have long working distance objectives in order to focus on the sample while looking through the stamp.
- Temperature control. A heater and thermometer are enclosed in the sample stage to control the temperature. The temperature control is used to change the properties of the PC film during the stacking procedure.
- Binoculars to search for flakes and follow the stacking procedure.
- Aperture diaphragm control lever. During most of the stacking process the sample will be focused through the glass slide having the PDMS/PC stamp and it will also be very important to be able to properly focus on the stamp itself. This will highly affect the resolution of the microscope, making it very hard to focus correctly. In order to avoid this issue, all the time the light goes through the glass slide the aperture-control feature of the microscope is used. By closing the aperture, the depth of focus increases, allowing to properly focus on the sample through the stamp and/or properly see the flakes on the stamp. This setting can also be used to highlight defects in flakes, as shown in Fig. S9.
- Camera and imaging software. Most of the stacking is done with the image acquired by the camera and using the imaging software of the microscope. This is especially important for the stacking procedure for three main reasons. Firstly, changing the contrast and colour saturation of the camera will allow to see defects in flakes which are not visible with the naked eye. Secondly, it allows to outline the shapes of the different flakes. For double gated stacks this is very important for the last step, in which the top and bottom gates will be aligned. As more layers are picked up with the PC stamp, visibility through the layers decreases. That means that depending on the thickness of the chosen graphite gate, and mainly hBN flakes, it is possible that when picking up the bottom graphite gate, the top graphite gate is no longer visible. In this case having outlined where the top gate is positioned with respect to the other flakes will be essential to have a proper alignment with the bottom gate. The same applies to the graphene, which will not be visible once the bottom hBN is picked up. Finally, saving the images of the stacking process is necessary to design the device design for the nanofabrication. Since the graphene and the gates may not be visible once the stack is finally dropped on a $SiO_2$ chip, having pictures with the position of the different flakes is essential to design the final device.
- Fan. A fan is added to cool down the stage faster. This can be useful after dropping the stack for example, since the stage will be at 180 °C.
- Filters. Using filters can help detect defects in the flakes which are not visible otherwise.
- White light source.

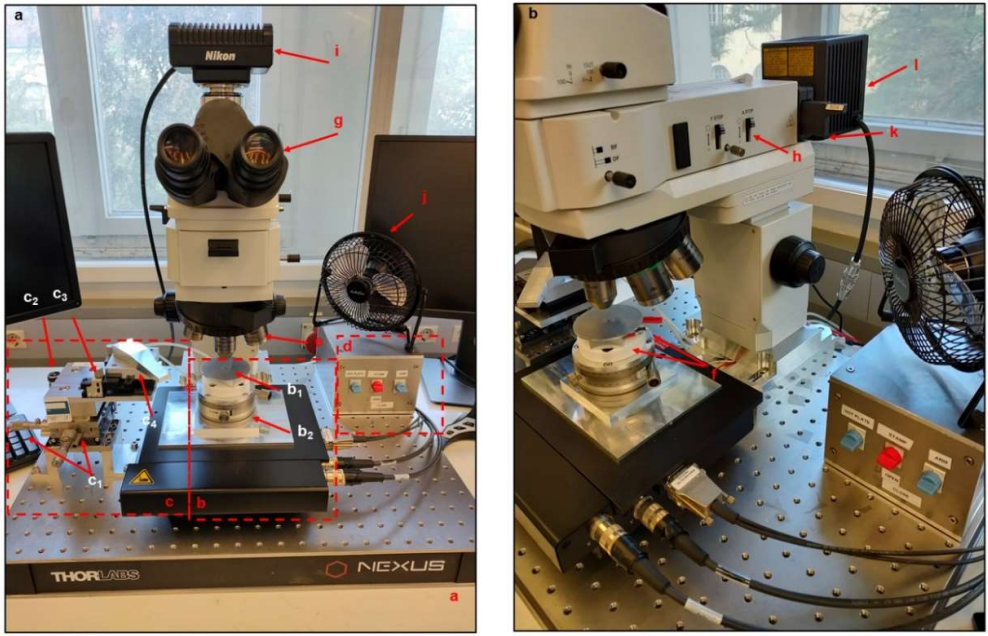

**Fig. S8. Transfer stage from the front and side view.** The letters correspond to the different parts explained above.

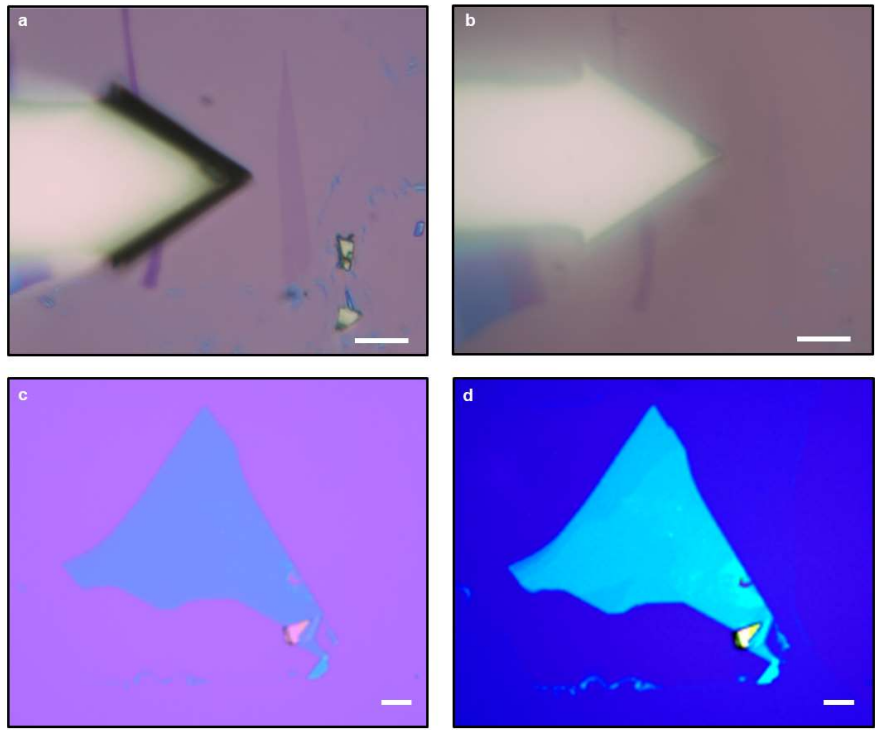

**Fig. S9. Optical image upon closing the aperture of the microscope. a,** The aperture diaphragm is closed, allowing to focus on the graphene flake and seeing clearly the AFM tip during cutting. **b,** The aperture diaphragm is open. The graphene flake is now barely visible. **c,** Apparent clean hBN flake. **d,** Upon closing the aperture and changing the colors and contrast one can clearly see defects in the hBN flake. Scale bar is 10 μm in all figures.

## F. Locking the twist-angle of MATBG by anchoring to the hBN edges

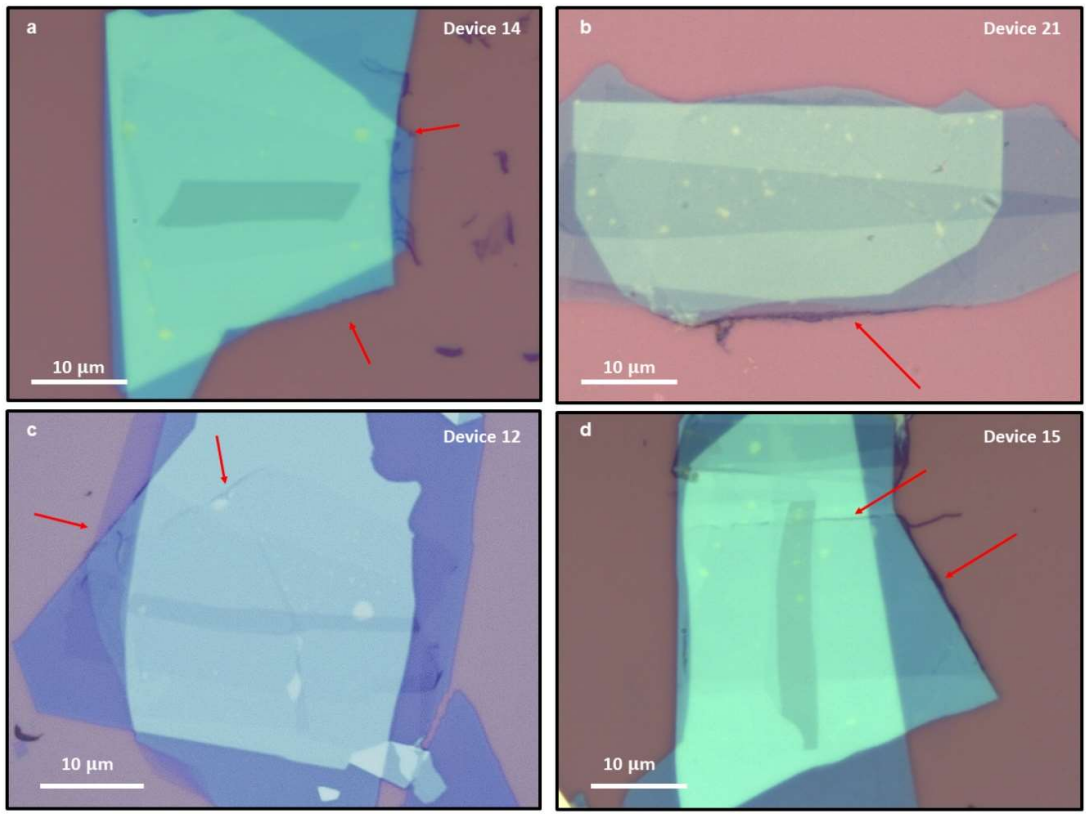

**Fig. S10. Optical images of 4 finalized stacks in which the clamping is particularly clear.** The clamping shows as an increased roughness on the edge of the hBN/graphene interface, signaled by the red arrows.

## G. Bubble statistics and device homogeneity

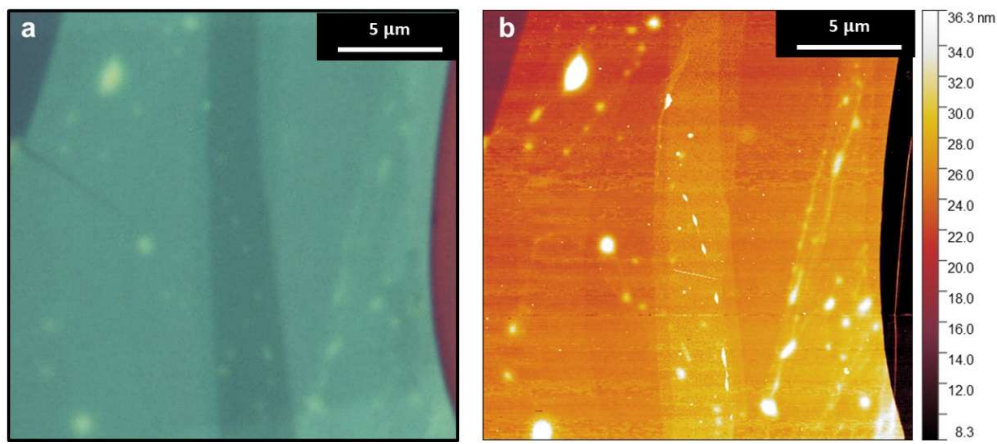

**Fig. S11. Comparison between an optical image with enhanced contrast and its AFM counterpart.** We observe no sharp difference in the presence of bubbles between the two images.

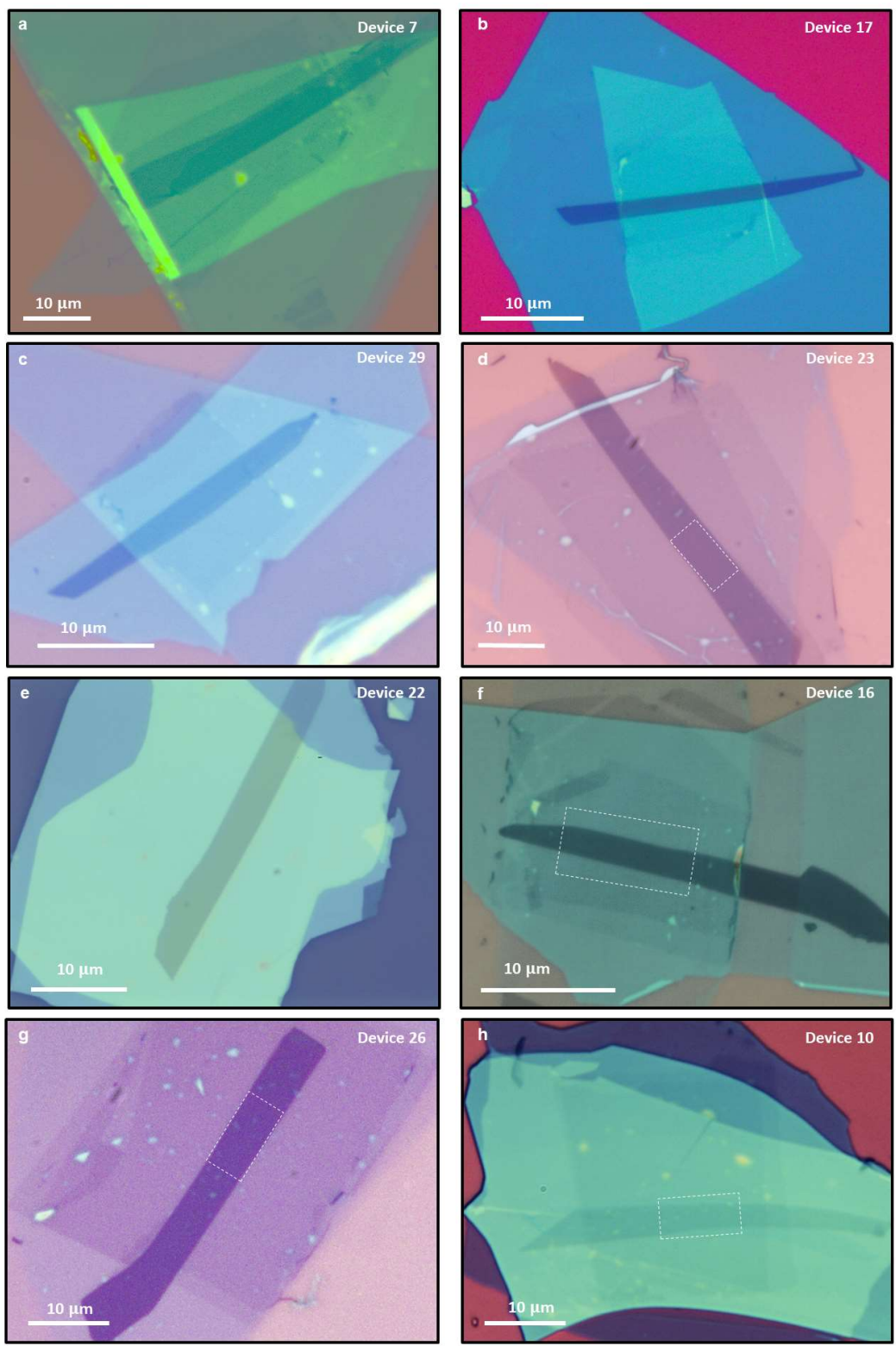

**Fig. S13. Several finalized stacks made with the described protocol. a-f,** Examples of stacks with the largest bubble free areas, with several of them having clean areas larger than 200 μm². **g, h,** Stacks which have small bubbles in the device area (inside the graphite gate). The dashed rectangles mark the 10 x 5 μm² area used to count the bubbles.

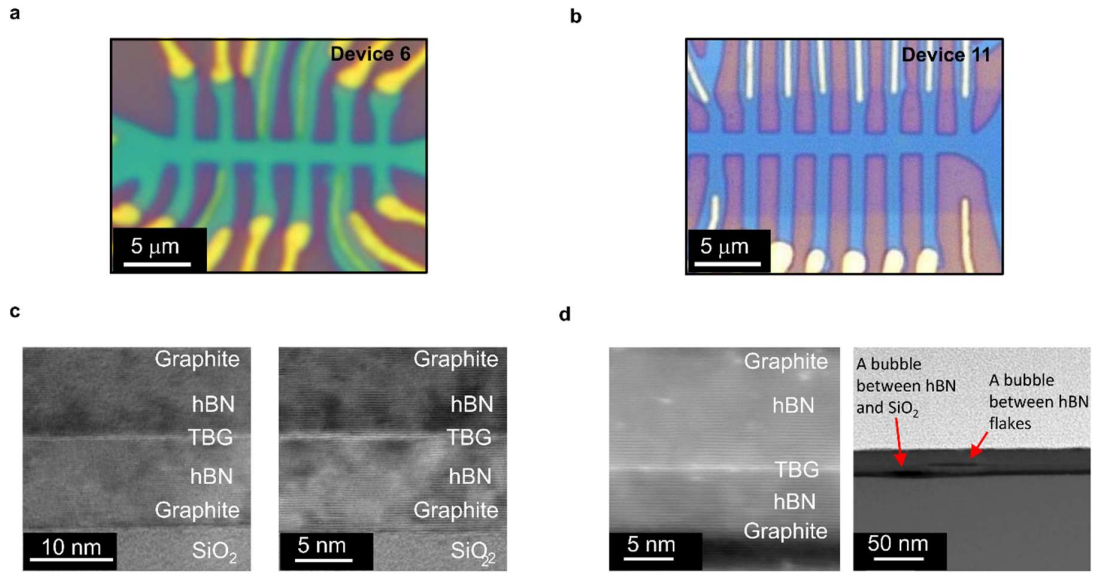

**Fig S12. Cross sectional STEM imaging of two devices fabricated using our protocol. a-b,** Optical images of two double graphite gated twisted graphene devices fabricated following the protocol described in the main text. **c-d,** ADF STEM images, taken in different regions of the device. The lattice fringes of the double-layer graphene sandwiched between the hBN layers, as well the graphite electrodes, are clearly visible in the STEM images. In most of the device area, the graphene/graphene and graphene/hBN, as well as hBN/graphite, interfaces are homogeneous and clean at the atomic scale. A few bubbles with a dimension of ~ 50 nm at hBN/substrate and hBN/graphene interfaces were visible in the second device, as shown in the bottom panel of **d**.

**H. Room *T* measurement of a large angle twisted bilayer graphene device**

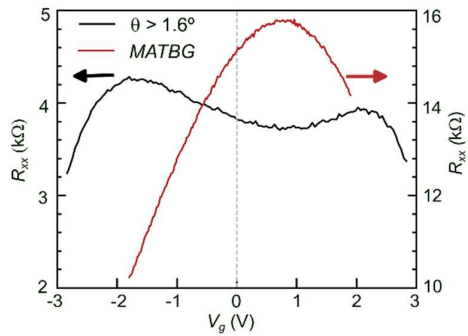

**Fig. S14. Transport of large-angle TBG device at room *T*.** Room *T* measurement comparing a large twist angle bilayer graphene ($\theta \gtrsim 1.6°$) and a magic-angle twisted bilayer graphene (MATBG) device with twist angle $\theta = 1.06 \pm 0.02°$, shown in the black and red curves, respectively.

# I. List of characterized devices

Here we summarize the information about the devices used for the characterization of the protocol. We state the twist angle, the density of bubbles and the calculated device area with angle homogeneity $\Delta\theta < 0.02°$. For the devices in the range $\theta = 1.1° \pm 0.1°$, we also write down whether superconductivity (SC) or a correlated insulator state (CI) was observed.

| Device | Twist Angle $\theta$ (°) | Bubbles >1μm² per 10x5 μm² | Bubbles <1μm² per 10x5 μm² | Nr. of consecutive contacts with $\Delta\theta < 0.02°$ | Distance between contacts | Width of Hall bar | Area with $\Delta\theta < 0.02°$ |
|---|---|---|---|---|---|---|---|
| 1 | 1.24 | 0 | 0 | 3 | 1.5 | 2 | 9 |
| 2 | 0 | 1 | 5 | - | - | - | - |
| 3 | 1.06 | 0 | 0 | 2 | 1.5 | 1.2 | 3.6 |
| 4 | 0.95 | 4 | 5 | 1 | 2 | 1.2 | 2.4 |
| 5 | 0 | 0 | 4 | - | - | - | - |
| 6 | 0 | 1 | 2 | - | - | - | - |
| 7 | 0 | 0 | 0 | - | - | - | - |
| 8 | 1.14 | 1 | 6 | 1 | 2 | 1.2 | 2.4 |
| 9 | 0 | 1 | 1 | - | - | - | - |
| 10 | 1.07 | 1 | 7 | 2 | 2 | 1 | 4 |
| 11 | 1.14 | 2 | 5 | 3 | 2 | 1 | 6 |
| 12 | 0.95 | 0 | 0 | 1 | 2 | 1.5 | 3 |
| 13 | 1.05 | 0 | 1 | 2 | 2 | 2 | 8 |
| 14 | 1.12 | 1 | 2 | 3 | 2 | 1.5 | 9 |
| 15 | 1.14 | 0 | 2 | 2 | 1.8 | 2 | 7.2 |
| 16 | 1.21 | 0 | 2 | 1 | 2 | 2 | 4 |
| 17 | 0 | 0 | 1 | - | - | - | - |
| 18 | 0.75 | 0 | 2 | - | - | - | - |
| 19 | 0 | 1 | 3 | - | - | - | - |
| 20 | 0.72 | 1 | 4 | - | - | - | - |
| 21 | 1.06 | 1 | 4 | 1 | 3 | 3 | 7.5 |
| 22 | 1.02 | 0 | 0 | 1 | 3.5 | 3 | 10.5 |
| 23 | 0.96 | 0 | 1 | 1 | 2 | 2 | 5 |
| 24 | 1.10 | 0 | 0 | 4 | 3 | 3 | 36 |
| 25 | 1.05 | 0 | 1 | 1 | 4.5 | 4.5 | 20 |
| 26 | 1.13 | 1 | 4 | 3 | 3 | 3 | 27 |
| 27 | 1.45 | 2 | 3 | - | - | - | - |
| 28 | 0.6 | 0 | 2 | - | - | - | - |
| 29 | 0 | 0 | 0 | - | - | - | - |
| 30 | 0 | 1 | 4 | - | - | - | - |
| 31 | 0.6 | 0 | 6 | - | - | - | - |
| 32 | 1.11 | 2 | 0 | 2 | 4 | 2 | 8 |
| 33 | 1.4 | 1 | 3 | - | - | - | - |
| 34 | 0 | 1 | 4 | - | - | - | - |

| Device | Twist Angle $\theta$ (°) | SC | $T_c$ 50% (K) | CI at $\nu$= -2 | CI at $\nu$= +2 |
|---|---|---|---|---|---|
| 3 | 1.06 | Yes | 2.88 | Yes | Yes |
| 8 | 1.14 | Yes | 1.45 | Yes | Yes |
| 10 | 1.07 | Yes | 1.07 | Yes | Yes |
| 11 | 1.14 | No | - | Yes | Yes |
| 13 | 1.05 | Yes | 0.35 | No | No |
| 14 | 1.12 | Yes | 2.41 | Yes | Yes |
| 15 | 1.14 | Yes | 2.72 | Yes | Yes |
| 21 | 1.06 | No | - | No | Yes |
| 22 | 1.02 | No | - | Yes | Yes |
| 24 | 1.10 | Yes | 2.16 | No | Yes |
| 25 | 1.05 | No | - | Yes | Yes |
| 26 | 1.13 | Yes | 0.44 | Yes | Yes |
| 32 | 1.11 | Yes | 1.62 | Yes | Yes |